\documentclass[letter,11pt]{article}
\usepackage[top=1in,bottom=1in,left=.8in,right=.8in]{geometry}
\usepackage{slashed}
\usepackage{epsfig}
\usepackage{comment}
\usepackage{cancel}
\usepackage{bbm}
\usepackage{array}
\usepackage{bigints}
\usepackage{booktabs}
\usepackage{color}
\usepackage{dsfont}
\usepackage{float}
\usepackage{framed}
\usepackage{graphicx}
\usepackage{indentfirst}
\usepackage{mathrsfs}
\usepackage{multirow}
\usepackage{simpler-wick}
\usepackage{subdepth}
\usepackage{titlesec}
\usepackage[dotinlabels]{titletoc}
\usepackage{wrapfig}
\usepackage[all]{xy}
\usepackage[vcentermath]{youngtab}
\usepackage{relsize}
\usepackage{hyperref}
\usepackage{xcolor}
\usepackage{soul}
\numberwithin{equation}{section}

\usepackage[utf8]{inputenc}
\usepackage{slashed}
\usepackage{amsmath}
\usepackage{amsfonts}
\usepackage{amssymb}
\usepackage{cite}
\usepackage{mathtools}

\usepackage{cleveref}
\crefname{section}{§}{§§}
\crefname{section}{§}{§§}

   \makeatletter
  \let\over=\@@over \let\overwithdelims=\@@overwithdelims
  \let\atop=\@@atop \let\atopwithdelims=\@@atopwithdelims
  \let\above=\@@above \let\abovewithdelims=\@@abovewithdelims
\renewcommand\section{\@startsection {section}{1}{\z@}%
{-3.5ex \@plus -1ex \@minus -.2ex}
{2.3ex \@plus.2ex}%
{\normalfont\large\bfseries}}

\renewcommand\subsection{\@startsection{subsection}{2}{\z@}%
{-3.25ex\@plus -1ex \@minus -.2ex}%
{1.5ex \@plus .2ex}%
{\normalfont\bfseries}}

\makeatletter
\DeclareFontFamily{OMX}{MnSymbolE}{}
\DeclareSymbolFont{MnLargeSymbols}{OMX}{MnSymbolE}{m}{n}
\SetSymbolFont{MnLargeSymbols}{bold}{OMX}{MnSymbolE}{b}{n}
\DeclareFontShape{OMX}{MnSymbolE}{m}{n}{
    <-6>  MnSymbolE5
   <6-7>  MnSymbolE6
   <7-8>  MnSymbolE7
   <8-9>  MnSymbolE8
   <9-10> MnSymbolE9
  <10-12> MnSymbolE10
  <12->   MnSymbolE12
}{}
\DeclareFontShape{OMX}{MnSymbolE}{b}{n}{
    <-6>  MnSymbolE-Bold5
   <6-7>  MnSymbolE-Bold6
   <7-8>  MnSymbolE-Bold7
   <8-9>  MnSymbolE-Bold8
   <9-10> MnSymbolE-Bold9
  <10-12> MnSymbolE-Bold10
  <12->   MnSymbolE-Bold12
}{}

\let\llangle\@undefined
\let\rrangle\@undefined
\DeclareMathDelimiter{\llangle}{\mathopen}%
                     {MnLargeSymbols}{'164}{MnLargeSymbols}{'164}
\DeclareMathDelimiter{\rrangle}{\mathclose}%
                     {MnLargeSymbols}{'171}{MnLargeSymbols}{'171}
\makeatother

\newcommand{\D}{\partial}

\newcommand{\LP}{\left(}
\newcommand{\RP}{\right)}
\newcommand{\KET}[1]{\left| #1 \right\rangle }

\newcommand{\mc}[1]{\mathcal{#1}}

\newcommand{\hb}{\bar{h}}

\begin{document}
\begin{titlepage}
\unitlength = 1mm
\ \\
\vskip 3cm
\begin{center}

{\LARGE{\textsc{Multi-particle Celestial Operator Product \\[1 ex]Expansions from the Boundary}}}

\vspace{1.25cm}
Mathew Calkins and Monica Pate

\vspace{.5cm}

{\it The Center for Cosmology and Particle Physics, New York University, New York, NY 10003}\\ 

\vspace{0.8cm}

\begin{abstract} 

In celestial holography, scattering particles in four-dimensional asymptotically flat spacetimes are dual to conformal primary field operators on the celestial sphere.  Multi-particle celestial operators can be formed from regularized coincident limits of single-particle celestial operators.  The singular terms in the operator product expansion of multi-particle operators are shown to be determined entirely by the singular terms in the operator product expansion between single-particle celestial operators, as expected in a standard conformal field theory. Boundary operator product expansions in celestial holography are known to be dual to subtle collinear limits of bulk scattering amplitudes.  The multi-particle operator product expansions derived from standard conformal-field theoretic techniques are shown to reproduce precisely the results from the corresponding bulk collinear limits in tree-level Yang-Mills and Einstein gravity. Finally, the coefficients of multi-particle celestial operator product expansions are derived from a third complementary method  that enforces bulk four-dimensional translational invariance as a global symmetry of the celestial dual. The results of all three methods agree precisely.

\end{abstract}

\vspace{1.0cm}
\end{center}

\end{titlepage}

\pagestyle{empty}
\pagestyle{plain}

\def\vx{{\vec x}}
\def\p{\partial}
\def\po{$\cal P_O$}
\def\i{{\rm initial}}
\def\f{{\rm final}}

\pagenumbering{arabic}
 

\tableofcontents

\section{Introduction}

The holographic principle --- the conjecture that systems of quantum gravity admit  alternative but entirely equivalent formulations as ordinary, non-gravitational systems in fewer spacetime dimensions --- is a rather profound statement about the nature of the physical world.  In particular, it implies that, under a suitable projection, the physics of quantum-gravitational systems manages to organize into that of a sensible lower-dimensional physical system with local interactions.  In the context of asymptotically Anti-de Sitter spacetimes, explicit top-down constructions from string theory provide compelling and concrete evidence for this perhaps surprising emergence of lower-dimensional physics \cite{Maldacena:1997re}.  On the other hand, a number of recent investigations in asymptotically flat spacetimes, specifically those pertaining to the recently-formulated celestial holographic proposal, are motivated from bottom-up arguments in which the justification for a local lower-dimensional dual is less clear.\footnote{Remarkably in AdS/CFT, a notion of locality in the form of an operator product expansion can be justified from purely bottom-up arguments \cite{KaplanLectureNotes}. It would be very encouraging if a similar statement could be made in celestial holography. See \cite{Crawley:2021ivb} for a somewhat related study.}  Beyond providing novel insight into the fundamental mechanics of quantum gravity, conventional local dual systems are also of practical use. Namely, if the dual descriptions of quantum gravitational systems are governed by ordinary and predictive laws of physics, then new lessons about quantum gravity may be accessed readily. 

According to the celestial holographic proposal, scattering in asymptotically flat spacetimes is holographically dual to a conformal field theory living on the celestial sphere.  Lorentz symmetry ${\rm SO}(1,3)$ in the bulk manifests as a global conformal symmetry ${\rm SL}(2, \mathbb{C})$ on the boundary. As a result, scattering particles in highest-weight representations of ${\rm SL}(2, \mathbb{C})$ transform like primary field operators in a 2D CFT under bulk Lorentz transformations, or equivalently boundary conformal transformations.  Accordingly, scattering amplitudes of these particles transform like correlation functions of primary operators in a 2D CFT. From this perspective, the positions of operators on the boundary in essence emerge from representation theory.  This makes it not at all obvious why they should behave like local operators, despite being labeled by a definite position.\footnote{Indeed there exists an explicit, loop-exact example of non-local behavior of a massive particle on the celestial sphere \cite{Himwich:2023njb}.} 

Remarkably, ${\rm SL} (2, \mathbb{C})$ highest-weight scattering states of massless particles formed from Mellin transformations of standard momentum eigenstates indeed behave like local operators on the celestial sphere. Specifically, their correlation functions are known to admit singularities in the coincident limit of two operator positions \cite{Strominger:2013lka,He:2015zea,Fan:2019emx,Pate:2019lpp}.  In a typical conformal field theory, this limit is governed by an operator product expansion (OPE) and is characteristic of a local field theory. The origin of these singularities in celestial holography can be traced back to collinear singularities of the corresponding scattering amplitudes between particles in plane wave states of definite momentum. Specifically, scattering amplitudes are known to admit universal singular behavior in the limit in which the external momenta of two massless particles are taken collinear. 

In conventional conformal field theories, the operator product expansion is a powerful organizing principle, allowing higher-point correlation functions to be expressed in terms of lower point correlation functions and ultimately reducing the minimal data needed to specify a CFT to the OPE coefficients and conformal dimensions of primary operators. This underlying organization is intimately related to the singularity structure of the correlation functions. It is thus of significant interest to determine the precise ways that holographic \emph{celestial} conformal field theories (\textit{i.e.}~those dual to quantum gravity in asymptotically flat spacetimes) admit conventional behavior. 

One relevant line of inquiry beyond the simple collinear limits described above, is the investigation into multi-collinear limits or in other words coincident limits involving more than two asymptotic particles. In particular, there is at least a naive discrepancy between the recursive structure of CFT correlation functions and QFT scattering amplitudes. Namely, the operator product expansion in conformal field theory always replaces a pair of operators with a single local operator, thereby relating $n$-point correlations to $(n-1)$-point correlations.  On the other hand, scattering amplitudes admit a multitude of different factorization channels as different intermediate particles are taken on-shell.  A key insight for bridging these two limits was first articulated in \cite{Ebert:2020nqf} who observed that the maximally singular limits of amplitudes that recursively probe all possible factorization channels are readily interpretable as successive celestial operator product expansions.  Note however, despite this initial encouraging observation, the general kinematics of three-particle factorization channels were soon after shown to be responsible for subtle non-local behavior of generalized celestial currents that ultimately leads to a breakdown of the Jacobi identity \cite{Ball:2022bgg}. Subsequently in \cite{Ball:2023sdz}, three- (as opposed to two-) particle factorization channels were explicitly worked out and reinterpreted as celestial operator product expansions. See also \cite{Magnea:2025zut} for a more recent treatment in non-abelian gauge theory.  Finally, in \cite{Guevara:2024ixn}, operator product expansions for certain ``multi-particle'' operators formed from regularized coincident limits of pairs of celestial gluons were determined from a careful analysis of factorization channels and collinear limits of bulk gluon scattering amplitudes in tree-level Yang-Mills.  Other related work includes various investigations into the role of soft current algebra descendants in single-particle celestial operator product expansions and constraints on celestial amplitudes \cite{Ebert:2020nqf,Banerjee:2020kaa,Banerjee:2020zlg, Banerjee:2020vnt,Banerjee:2021cly,Hu:2021lrx, Banerjee:2021dlm,Banerjee:2023zip,Banerjee:2023bni,Banerjee:2025grp,Adamo:2022wjo, Ren:2023trv}, the analysis of representation theoretic aspects of multi-particle celestial operators \cite{Kulp:2024scx}, and the recent identification of certain composites of celestial operators associated to bulk asymptotic symmetries as marginal operators in celestial CFT \cite{Imseis:2025awd}.

Our work builds upon and extends these results, particularly those in \cite{Guevara:2024ixn}. Specifically, the regularization used to form the ``multi-particle'' operators in \cite{Guevara:2024ixn} is a standard prescription in the 2D CFT literature. Notably, in a conventional 2D CFT, the singular terms in the OPEs of these ``composite'' operators are directly determined by the singular terms in the OPEs of the constituents.  In this work, we apply standard 2D CFT techniques to determine the OPEs of multi-particle operators. We find precise agreement with \cite{Guevara:2024ixn} for gluons and extend their results to any type of massless spinning particle.  Moreover, our method enables us to fix the OPE coefficients of multi-particle operators appearing in the multi-particle OPE, which were not fully determined by the analysis of \cite{Guevara:2024ixn}, as well as access singularities that arise from mixed holomorphic and anti-holomorphic limits.  While this boundary method is guaranteed to be correct by complex analysis, which we demonstrate explicitly, we also reproduce these results by two other independent methods. In particular, we explain how to derive the result from a corresponding bulk collinear limit, here again generalizing the procedures presented in \cite{Guevara:2024ixn} to Einstein gravity at tree-level.  Finally, we show perfect agreement with a third complementary approach that enforces 4D bulk Poincar\'e symmetry as a global symmetry on the boundary. This last method generalizes the analysis in \cite{Pate:2019lpp, Himwich:2021dau} to multi-particle operators and their OPEs. 

Our results thus demonstrate that the OPE coefficients of multi-particle operators are not independent data that must be specified \emph{in addition to}  the OPE coefficients of the single-particle operators, but rather are \emph{entirely determined by} the OPE coefficients of the single-particle operators.  Constraints of this type that reduce the number of independent parameters are crucial for the general tractability of the celestial holography program.  Note however, since the derivation of singular terms in the multi-particle OPE only exploits an underlying mathematical identity, unfortunately this analysis cannot be used to evaluate the consistency of the singularity structure of celestial amplitudes with that of an underlying 2D field theory. In this context, the revealed minimal input, namely the singular terms in the single-particle OPEs, are well-studied and known to be derivable from a number of different complementary methods \cite{Fan:2019emx,Pate:2019lpp,Himwich:2021dau,Jiang:2021ovh,Jiang:2021csc,Adamo:2021zpw,Bu:2021avc,Monteiro:2022lwm,Guevara:2022qnm}. It also worth emphasizing that in this context the standard CFT method applies directly to the mixed helicity cases, so the boundary method in fact determines both holomorphic and anti-holomorphic limits simultaneously and can access singularities that only arise from mixed limits. Prior to this work, the holomorphic and anti-holomorphic singularities could only be calculated by combining the results from separate holomorphic and anti-holomorphic collinear limits in the bulk.  Likewise, the symmetry method is similarly restricted to account properly for either the structure of the holomorphic or anti-holomorphic global conformal descendants.   

Throughout this work, the explicit expressions we derive for celestial operator products expansions pertain strictly to tree-level scattering. However, we expect the 2D CFT technique based on a mathematical identity will extend to loop-level,\footnote{Apart from subtleties that may arise due to branch cuts.} as will the symmetry-based method, provided that loop-corrected expressions for single-particle celestial operators are used as input.\footnote{Such as the 1-loop corrections to celestial gluon OPEs  \cite{Bhardwaj:2022anh,Krishna:2023ukw,Bhardwaj:2024wld} or the loop corrections arising from twistor anomalies that play a crucial role in associativity \cite{Costello:2022wso,Costello:2022upu,Bittleston:2022jeq,Fernandez:2023abp,Zeng:2023qqp,Fernandez:2024qnu,Serrani:2025oaw}.} 

This paper is organized as follows.  In section \ref{sec:boundary-derivation}, we use standard conformal field-theoretic techniques to derive multi-particle celestial operator product expansions from single-particle operator product expansions.  The mathematical identity guaranteeing the general applicability of this method is presented in appendix \ref{app:explain-Wick}.  Then, in subsection \ref{subsec:positive-helicity-gravitons}, we explicitly perform the analysis for operator product expansions involving only positive helicity gravitons. We use this specialized context to identify and clarify points of the analysis that are slightly atypical but straightforwardly generalize from a standard conformal field theory. Then, in subsection \ref{subsec:holomorphic-allspin}, we extend our results from the previous section to recover all holomorphic singular terms in operator product expansions of general massless particles involving any combination of helicities. When restricted to gluons coupled by the Yang-Mills three-point interaction, our results are consistent with those in \cite{Guevara:2024ixn}. Finally, we readily obtain all (\textit{i.e.}~both holomorphic and anti-holomorphic) singular terms by applying the technique to the full single-particle operator product expansion involving both holomorphic and anti-holomorphic singularities.  Our final result includes additional contributions beyond those that arise from a simple conjugation of the result in subsection \ref{subsec:holomorphic-allspin}, due to overlapping holomorphic and anti-holomorphic singularities. We derive this most general OPE in subsection \ref{subsec:general-ope}.  

In section \ref{sec:bulk-collinear-limits}, we turn to a bulk derivation of multi-particle operator product expansions from collinear limits of scattering amplitudes. In subsection \ref{subsec:bulk-single-particle}, we derive the single-graviton contributions to graviton multi-particle OPEs for all possible combinations of helicities and find perfect agreement with our boundary calculations in section \ref{sec:boundary-derivation}.  In subsection \ref{subsec:bulk-multi-particle}, we restrict our attention to maximally helicity-violating (MHV) graviton amplitudes and derive multi-particle graviton OPEs, including both single-graviton and multi-graviton contributions.  Again, we find perfect agreement with our results in section \ref{sec:boundary-derivation}. The spinor helicity conventions used in this section can be found in appendix \ref{sec:conventions}.

In the penultimate section \ref{sec:symmetry}, we present yet a third method for deriving multi-particle celestial operator product expansions:~enforcing bulk symmetry constraints. In subsections \ref{subsec:sl2c} and \ref{subsec:translations}, we use a similar technique as in section \ref{sec:boundary-derivation} to derive the transformation laws of multi-particle celestial operators under Poincar\'e from the known Poincar\'e transformation laws of single-particle celestial operators.  Analogues of these transformations in the Carrollian basis can be found in \cite{Kulp:2024scx}. Then in subsection \ref{subsec:symmetry-single-particle} and \ref{subsec:symmetry-multi-particle}, we show how invariance of the operator product expansion under Poincar\'e, especially translational symmetry, places constraints on the operator product expansion coefficients and we find that the coefficients derived in section \ref{sec:boundary-derivation} and \ref{sec:bulk-collinear-limits} respect these constraints. Subsection \ref{subsec:symmetry-single-particle} concerns the OPE coefficients of single-particle contributions to the multi-particle OPE and subsection \ref{subsec:symmetry-multi-particle} treats the OPE coefficients of multi-particle contributions. In appendix \ref{app:singleParticleRecursion}, we derive the solution to the system of constraints implied by the symmetry analysis in subsection \ref{subsec:symmetry-single-particle}. 

In the final section \ref{sec:discussion}, we summarize our results, place them in the broader context of the celestial holography program and discuss directions for future research that are enabled by this work.

\section{Multi-particle OPEs from single-particle OPEs}
\label{sec:boundary-derivation}

The usual celestial dictionary \cite{Pasterski:2017kqt} relates bulk scattering amplitudes in asymptotically flat 4D spacetime to correlation functions in a 2D boundary conformal field theory.  Specifically, bulk massless single-particle asymptotic states of definite boost weight, formed by Mellin-transforming momentum eigenstates with respect to energy, are holographically dual to primary field operators on the 2D boundary.  Universal collinear limits of scattering amplitudes, in which a pair of external momenta are taken to be asymptotically collinear, are realized in boundary correlation functions by coincident limits of the corresponding celestial operators.  In conventional conformal field theories, the coincident limit of operators is governed by the operator product expansion, which manifests in the singularity structure of correlation functions. In this section, we investigate the corresponding singularities in celestial amplitudes.

More precisely, we study operator product expansions involving multi-particle operators formed by the following regularized product of single-particle celestial operators
\begin{equation} \label{def-normal-ordering}
    \begin{split}
        :\mathcal{O}_{1} \mathcal{O}_{2}: (z_2, \bar{z}_2) \equiv \oint_{z_2} \frac{dz_1}{2 \pi i} \oint_{\bar{z}_2} \frac{d\bar{z}_1}{2 \pi i} \frac{1}{z_{12}\bar{z}_{12}}\mathcal{O}_{1}(z_1, \bar{z}_1)\mathcal{O}_{2}(z_2, \bar{z}_2).
    \end{split}
\end{equation}
Note, the contour integral in \eqref{def-normal-ordering} picks out the non-singular and non-vanishing part of the OPE in the limit $(z_1, \bar{z}_1) \to (z_2, \bar{z}_2)$.  Explicitly in the conventions of \cite{DiFrancesco:1997nk}, given two local operators $\mc{O}_1, \mc{O}_2$ with OPE 
\begin{equation*}
\mc{O}_1(z_1, \bar{z}_1) \mc{O}_2(z_2, \bar{z}_2) = \sum_{n, \bar{n} \in \mathbb{Z}} z_{12}^n \bar{z}_{12}^{\bar{n}} (\mc{O}_1 \mc{O}_2)_{n, \bar{n}}(z_2, \bar{z}_2),
\end{equation*}
the normal-ordered product or composite operator $:\mc{O}_1 \mc{O}_2:$ is the coefficient at $n = \bar{n} = 0$.
Here and throughout, we use double subscripts to denote coordinate differences $z_{ij} \equiv z_i - z_j$ and $\bar{z}_{ij} \equiv \bar{z}_i - \bar{z}_j$ and the following big-O notation. A multi-variable function $f$ is written as $f(z,\bar{z}) = g(z, \bar{z}) + \mc{O} \LP z^a \bar{z}^{\bar{a}} \RP$ when the multi-variable Laurent expansion of $f - g$ only contains non-vanishing terms $z^n \bar{z}^{\bar{n}}$ for $n \geq a$ and $\bar{n} \geq \bar{a}$. Likewise $\Theta \LP z^a \bar{z}^{\bar{a}} \RP$ is used to denote a series which only contains a $z^a \bar{z}^{\bar{a}}$ term.  For example, a general OPE decomposes in the following way:
\begin{equation} \label{OPE-decomp-explained}
\mc{O}_1(z_1, \bar{z}_1) \mc{O}_2(z_2, \bar{z}_2) = \underbrace{\wick{ \c {\mc{O}}_1(z_1, \bar{z}_1) \c {\mc{O}}_2(z_2, \bar{z}_2)}}_{\mbox{\scriptsize{singular as $1 \to 2$}}} + \overbrace{\underbrace{:\mc{O}_1 \mc{O}_2:(z_2, \bar{z}_2)}_{\mbox{\scriptsize{constant as $1 \to 2$}}} + \underbrace{\mc{O}\LP z_{12}^1 \bar{z}^0_{12}\RP + \mc{O} \LP z_{12}^0 \bar{z}_{12}^1\RP}_{\mbox{\scriptsize{vanishing as $1 \to 2$}}}}^{\mc{O}\LP z_{12}^0 \bar{z}_{12}^0 \RP \mbox{\scriptsize{, \textit{i.e.} regular as $1 \to 2$}}}.
\end{equation}

In our analysis, $\mc{O}_i$ are celestial operators, whose $n$-point correlation functions are equal to the Mellin-transform of scattering amplitudes of massless particles with respect to energy. In this work, we refer to composite operators formed from normal-ordered products of the single-particle celestial operators $\mc{O}_i$ according to \eqref{def-normal-ordering} as ``multi-particle'' operators.  Although it is tempting to identify our multi-particle celestial operators as the asymptotic states of bound bulk scattering particles, note that they differ subtly from standard bound states in quantum field theory.  More precisely, scattering of bound states in quantum field theory is extracted via the LSZ procedure as the \emph{residue} of poles in Fourier transforms of time-ordered correlation functions.  Although the regularization of poles in \eqref{def-normal-ordering} is reminiscent of the LSZ procedure, note that here we remove the poles by \emph{subtraction} as opposed to extraction of a residue.  

Initial investigations into operator product expansions involving multi-gluon celestial operators of the form \eqref{def-normal-ordering} were carried out in \cite{Guevara:2024ixn} where the authors determined the OPEs from a careful analysis of holomorphic collinear limits of bulk scattering amplitudes.  In this section, we introduce a pure boundary approach that exploits standard CFT methods to compute the OPE of composite operators \eqref{def-normal-ordering} from OPEs of the constituents.  We exploit the fact that the regularization appearing in \eqref{def-normal-ordering} is a standard prescription for forming composite operators in a conventional conformal field theory and that there is a simple and explicit procedure for determining the singular terms in OPEs involving these composites from the OPEs of the constituents \cite{DiFrancesco:1997nk}.  Specifically, using $\wick{\c{\mathcal{O}_1}(z_1, \bar{z}_1) \c {\mathcal{O}_2}(z_2, \bar{z}_2)}$ to denote the singular terms in the OPE between $\mathcal{O}_1(z_1, \bar{z}_1)$ and $\mathcal{O}_2(z_2, \bar{z}_2)$,\footnote{Explicitly in the conventions of \cite{DiFrancesco:1997nk}, the Wick contraction $\wick{\c{\mc{O}_1} \c{\mc{O}_2}}$ is defined to be the singular part 
\begin{equation*}
\wick{ \c {\mc{O}}_1(z_1, \bar{z}_1) \c {\mc{O}}_2(z_2, \bar{z}_2)} = \sum_{\substack{n, \bar{n} \in \mathbb{Z} \\ \min(n, \bar{n}) < 0}} z_{12}^n \bar{z}_{12}^{\bar{n}} (\mc{O}_1 \mc{O}_2)_{n, \bar{n}} (z_2, \bar{z}_2).
\end{equation*}}
we have
\begin{equation} \label{generalizedwick-1}
    \begin{split}
\wick{ \c {\mc{O}}_1(z_1, \bar{z}_1) :\mc{O}_2 \c{\mc{O}}_3}:(z_3, \bar{z}_3) & = \oint_{z_3} \frac{dz_2}{2 \pi i } \frac{1}{z_{23}} \oint_{\bar{z}_3} \frac{d \bar{z}_2}{2 \pi i  }  \frac{1}{\bar{z}_{23}}\Bigg[ \wick{\c {\mc{O}_1}(z_1, \bar{z}_1) \c {\mc{O}_2}} (z_2, \bar{z}_2) \mc{O}_3(z_3, \bar{z}_3)\\
&  \quad \quad \quad \quad \quad  \quad \quad \quad \quad  \quad \quad \quad  + \mc{O}_2(z_2, \bar{z}_2) \wick{\c {\mc{O}_1}(z_1, \bar{z}_1)  \c {\mc{O}_3}}(z_3, \bar{z}_3) \Bigg].
    \end{split}
\end{equation}
This formula can be thought of as a generalization of Wick's theorem that is applicable to generic interacting conformal field theories.  A common application of the formula is to determine the OPEs between Virasoro descendants, for example the OPE $T(z):TT:(0)$.  To clarify the assumptions underlying this formula, in appendix \ref{app:explain-Wick} we show that \eqref{generalizedwick-1} holds with no reference to operator product expansions or other field-theoretic ideas.  Rather, it is a mathematical identity relating different singular and regular terms in various limits of a function of multiple complex variables. 

In this section, we apply this formula to the known single-particle celestial operator product expansions \cite{Himwich:2021dau} and, in cases where a comparison can be made, find precise agreement with the results in \cite{Guevara:2024ixn}. To illustrate minor subtleties, we begin with an explicit presentation of the all-positive-helicity graviton OPE in tree-level Einstein gravity. We then generalize the calculation and determine terms in multi-particle OPEs resulting from holomorphic singularities in single-particle OPEs between general combinations of massless spinning particles, with the holomorphic singularities in Einstein gravity and Yang-Mills as special cases. Finally, we present the result for generic helicity combinations, including all possible holomorphic and anti-holomorphic singularities.  This final result follows from an exact treatment of \eqref{generalizedwick-1}. 

\subsection{Positive helicity gravitons}
\label{subsec:positive-helicity-gravitons}
The celestial duals of asymptotically flat gravitational theories include conformal primaries $G^\pm_\Delta$, which are the boundary representations of bulk gravitons.  Here the superscript $\pm$ specifies the graviton's helicity and $\Delta$ is the graviton boost weight \cite{Pasterski:2017kqt}.  On the boundary, these define the operator's scaling dimension $\Delta$ and conformal spin $\pm 2$. In \cite{Pate:2019lpp}, it was found that positive-helicity graviton operators respect a leading-order tree-level OPE of the form 
\begin{equation} \label{G^+G^+-primary}
G_{\Delta_1}^+(z_1, \bar{z}_1) G_{\Delta_2}^+(z_2, \bar{z}_2) \sim - \frac{\kappa}{2} \frac{\bar{z}_{12}}{z_{12}} B(\Delta_1 - 1, \Delta_2 - 1) G^+_{\Delta_1 + \Delta_2}(z_2, \bar{z}_2),
\end{equation}
where $\kappa = \sqrt{32 \pi G}$, here ``$\sim$'' means equivalence modulo both descendants and regular terms, and $B(x,y) = \Gamma(x)\Gamma(y) /\Gamma(x+y)$ is the Euler beta function. The OPE coefficient can be either extracted from a collinear limit of bulk gravitational scattering or derived by enforcing asymptotic symmetries in the boundary theory \cite{Pate:2019lpp, Himwich:2021dau}.  
 
We now use \eqref{generalizedwick-1} to deduce the multi-particle OPE of the form $G^+_{\Delta_1}(z_1, \bar{z}_1) : G^+_{\Delta_2} G^+_{\Delta_3}:(z_3, \bar{z}_3)$.  The main analytic subtlety is that \eqref{generalizedwick-1} is sensitive to all singular holomorphic and anti-holomorphic terms in the $G^+G^+$ OPE, including those that arise from anti-holomorphic descendants.\footnote{This subtlety does not arise in the standard application to the $T:TT:$ OPE.  This is because $T$ is in a shortened representation without anti-holomorphic descendants as can be seen from the current conservation equation $\bar{\partial}T=0$.}  As a result, even the derivation of the leading contributions from global conformal primaries requires that we keep contributions from the first anti-holomorphic descendants at intermediate steps.  Here, we provide a careful explanation of the $G^+:G^+G^+:$ OPE calculation in Einstein gravity before presenting a more succinct derivation of the general case in the following subsection. 
 
As described above, we need the contribution from the first anti-holomorphic descendant. Since the OPE coefficients of descendants are fixed in terms of the OPE coefficients of primaries  by conformal symmetry, \eqref{G^+G^+-primary} readily extends to \cite{Himwich:2021dau}   
\begin{equation} \label{G^+-with-1desc}
\begin{aligned}
G^+_{\Delta_1} (z_1, \bar{z}_1) G^+_{\Delta_2} (z_2, \bar{z}_2)  & = -\frac{\kappa}{2}\frac{\bar{z}_{12}}{z_{12}} B(\Delta_1 - 1, \Delta_2 - 1) G^+_{\Delta_1 + \Delta_2}(z_2, \bar{z}_2)   -\frac{\kappa}{2} \frac{\bar{z}_{12}^2}{z_{12}} B(\Delta_1, \Delta_2 - 1) \bar{\partial} G_{\Delta_1 + \Delta_2}^+(z_2, \bar{z}_2) \\
\ & \quad \quad + \mbox{(subleading descendants)} + \mc{O} \LP z_{12}^0 \bar{z}_{12}^0 \RP.
\end{aligned}
\end{equation}
Differentiating with respect to $\bar{z}_1$ gives 
\begin{equation} \label{G+G+-withderiv}
\begin{aligned}
\left( \bar{\D} G^+_{\Delta_1} \right) (z_1, \bar{z}_1) G^+_{\Delta_2} (z_2, \bar{z}_2)  & =-\frac{\kappa}{2} \frac{1}{z_{12}} B(\Delta_1 {-} 1, \Delta_2 {-} 1) G^+_{\Delta_1 + \Delta_2}(z_2, \bar{z}_2)  - \kappa\frac{\bar{z}_{12}}{z_{12}} B(\Delta_1, \Delta_2 {-} 1) \bar{\partial} G_{\Delta_1 + \Delta_2}^+(z_2, \bar{z}_2) \\
\ & \quad \quad  + \mbox{(subleading descendants)} + \mc{O} \LP z_{12}^0 \bar{z}_{12}^0 \RP.
\end{aligned}
\end{equation}
Substituting \eqref{G^+-with-1desc} into \eqref{generalizedwick-1} and distributing terms gives 
\begin{equation} \label{G+G+-fullint}
\begin{split}
G_{\Delta_1}^+(z_1, \bar{z}_1) :G_{\Delta_2}^+ G_{\Delta_3}^+:(z_3, \bar{z}_3) & = -\frac{\kappa}{2}\oint_{z_3} \frac{dz_2}{2 \pi i z_{23}} \oint_{\bar{z}_3} \frac{d\bar{z}_2}{2 \pi i \bar{z}_{23}} \\
& \ \ \ \Bigg[ \frac{\bar{z}_{12}}{z_{12}} B(\Delta_1 - 1, \Delta_2 - 1) G^+_{\Delta_1 + \Delta_2}(z_2, \bar{z}_2) G_{\Delta_3}^+(z_3, \bar{z}_3) \\
\ & \ \ \ + \frac{\bar{z}_{12}^2}{z_{12}} B(\Delta_1, \Delta_2 - 1) \left( \bar{\D} G_{\Delta_1 + \Delta_2}^+ \right)(z_2, \bar{z}_2) G_{\Delta_3}^+(z_3, \bar{z}_3)  \\
\ & \ \ \ + \frac{\bar{z}_{13}}{z_{13}} B(\Delta_1 - 1, \Delta_3 - 1) G_{\Delta_2}^+(z_2, \bar{z}_2) G^+_{\Delta_1 + \Delta_3}(z_3, \bar{z}_3) \\
\ & \ \ \ +  \frac{\bar{z}_{13}^2}{z_{13}} B(\Delta_1, \Delta_3 - 1) G_{\Delta_2}^+(z_2, \bar{z}_2) \bar{\D} G_{\Delta_1 + \Delta_3}^+(z_3, \bar{z}_3)  \Bigg] \\
\ & \ \ \ + \mbox{(subleading descendants contributions)}  + \mc{O} \LP z_{13}^0 \bar{z}_{13}^0 \RP. 
\end{split}
\end{equation}

The third and fourth terms in square brackets are straightforward to evaluate; the integrals implement the $2 \to 3$ normal-ordered limit in \eqref{def-normal-ordering} and so give exclusively multi-particle contributions. The first term likewise gives only multi-particle terms, but requires slightly more work.  Decomposing $z_{12} = z_{13} - z_{23}$ and $\bar{z}_{12} = \bar{z}_{13} - \bar{z}_{23}$ and Taylor-expanding in $z_{23}$ and $\bar{z}_{23}$, the first line in square brackets involves the integral
\begin{equation} \label{G+G+-int1}
\sum_{n=0}^\infty \frac{1}{z_{13}^{n+1}} \oint_{z_3} \frac{dz_2}{2 \pi i z_{23}} \oint_{\bar{z}_3} \frac{d\bar{z}_2}{2 \pi i \bar{z}_{23}} (\bar{z}_{13} - \bar{z}_{23}) z_{23}^n \underbrace{G^+_{\Delta_1 + \Delta_2}(z_2, \bar{z}_2) G^+_{\Delta_3}(z_3, \bar{z}_3)}_{\sim \# \frac{\bar{z}_{23}}{z_{23}} G^+_{\Delta }(z_3, \bar{z}_3)  + \mathcal{O} \LP \bar{z}_{23}^2 z_{23}^{-1} \RP },
\end{equation}
where in the underbrace we provide the schematic form of the singular terms in the OPE. Since each singular term carries a positive power of $\bar{z}_{23}$, each is killed by the $\bar{z}_2$ integration, and only the normal-ordered product $:G^+_{\Delta_1 + \Delta_2} G^+_{\Delta_3}:$ arising from the $n = 0$ term in \eqref{G+G+-int1} survives.

The second term in \eqref{G+G+-fullint} is the most interesting and is the lone source of single-particle contributions to the composite OPE. Substituting \eqref{G+G+-withderiv}, including the leading regular term and expanding in $z_{23}$ and $\bar{z}_{23}$ as before, this becomes 
\begin{equation}
\begin{aligned}
& \sum_{n=0}^\infty \frac{1}{z_{13}^{n+1}} \oint_{z_3} \frac{dz_2}{2 \pi i z_{23}} \oint_{\bar{z}_3} \frac{d\bar{z}_2}{2 \pi i \bar{z}_{23}} (\bar{z}_{13} - \bar{z}_{23})^2 z_{23}^n \Bigg[-\frac{\kappa}{2} \frac{1}{z_{23}} B(\Delta_1 + \Delta_2 - 1, \Delta_3 - 1) G^+_{\Delta_1 + \Delta_2 + \Delta_3}(z_3, \bar{z}_3) \\
&  - \kappa  \frac{\bar{z}_{23}}{z_{23}} B(\Delta_1 + \Delta_2, \Delta_3 - 1) \bar{\partial} G_{\Delta_1 + \Delta_2 + \Delta_3}^+(z_3, \bar{z}_3) + : \left( \bar{\D}G^+_{\Delta_1 + \Delta_2} \right) G^+_{\Delta_3}: + \mc{O}\LP z_{23}^1 \bar{z}_{23}^0 \RP + \mc{O}\LP z_{23}^0 \bar{z}_{23}^1 \RP \Bigg].
\end{aligned}
\end{equation} 
We observe that the $n = 1$ term extracts the only single-particle contribution to the multi-particle OPE, namely $G^+_{\Delta_1 + \Delta_2 + \Delta_3}$. On the other hand, the $n=0$ term only gives multi-particle contributions, which are related to the previously-found multi-particle contributions by descendancy relations. Suppressing the latter and collecting results, we have finally
\begin{equation}\label{GGGplusOPE-final}
\begin{split} 
G_{\Delta_1}^+&(z_1, \bar{z}_1)~ {:}G_{\Delta_2}^+ G_{\Delta_3}^+{:}(z_3, \bar{z}_3) \\
&  \sim \frac{\kappa^2}{4}\frac{\bar{z}_{13}^2}{z^2_{13}} B(\Delta_1, \Delta_2 - 1) B(\Delta_1 + \Delta_2 - 1, \Delta_3 - 1) G^+_{\Delta_1 + \Delta_2 + \Delta_3}(z_3, \bar{z}_3) \\
\ & \ \ \ -\frac{\kappa}{2} \frac{\bar{z}_{13}}{z_{13}} B(\Delta_1 - 1, \Delta_2 - 1) {:}G^+_{\Delta_1 + \Delta_2} G_{\Delta_3}^+{:} (z_3, \bar{z}_3)  -\frac{\kappa}{2} \frac{\bar{z}_{13}}{z_{13}}  B(\Delta_1 - 1, \Delta_3 - 1) {:}G_{\Delta_2}^+ G^+_{\Delta_1 + \Delta_3}{:}(z_3, \bar{z}_3). 
\end{split}
\end{equation}
In the above expression, we display only the leading single-particle and composite contributions.  We later explain how to derive this result from a corresponding bulk calculation.

\subsection{``Pure'' holomorphic singularities for general helicity}
\label{subsec:holomorphic-allspin}

We now generalize the calculation above to generic interactions between massless spinning particles.  For clarity of presentation and direct comparison with bulk methods, here we focus on contributions to the multi-particle OPE that arise from only holomorphic singular terms in the single-particle OPEs. We lift this restriction in the next subsection.  

The tree-level holomorphic singular terms in operator product expansions between celestial primary operators representing massless spinning particles of general helicity were found in \cite{Himwich:2021dau} to take the universal form   
\begin{equation} \label{general-single-particle-OPE}
\mc{O}_1(z_1, \bar{z}_1) \mc{O}_2(z_2, \bar{z}_2) = \frac{1}{z_{12}} \sum_I \sum_{n=0}^\infty  \frac{\gamma_{s_I}^{s_1, s_2}}{n!} B( 2 \bar{h}_1 +p_{12I}+ n, 2 \bar{h}_2 +p_{12I} ) \bar{z}_{12}^{p_{12I}+ n} \bar{\D}^n \mc{O}_I (z_2,\bar{z}_2) + \mc{O}\big(z_{12}^0 \big). 
\end{equation}
Here $I$ indexes celestial primaries corresponding to single-particle asymptotic states, $s_i = h_i - \bar{h}_i$ is the conformal spin of the $i$th particle, and $p_{12I}$ is shorthand for the combination of spins
\begin{equation} \label{def-p}
    p_{12I} \equiv s_1+s_2-s_I-1. 
\end{equation}
In \cite{Pate:2019lpp, Himwich:2021dau}, it was found that the conformal spins were related to the dimension $d_V$ of the three-point interaction vertex mediating the process $12 \to I$ in the bulk theory by the constraint 
\begin{equation} \label{p-constraint}
    p_{12I} = d_V - 4,
\end{equation}
or equivalently  
\begin{equation} \label{spin-constraint}
    \begin{split}
        s_1+s_2-s_I =  d_V - 3 \geq 0.  
    \end{split}
\end{equation}
Finally it is also helpful to note that \eqref{general-single-particle-OPE}, \eqref{p-constraint} and \eqref{spin-constraint} together imply the fusion rule:
\begin{equation} \label{fusion-rule}
    \begin{split}
        \Delta_{I} = \Delta_1+\Delta_2+d_V-5, \quad \quad \quad 
        s_I = s_1+s_2-d_V+3.
    \end{split}
\end{equation} 
The coefficient  $\gamma_{s_I}^{s_1, s_2}$  vanishes for $s_1+s_2-s_I < 0$ but is otherwise is proportional by pure numbers to the corresponding coupling constant for a  bulk three-point interaction between particles of (outgoing) helicity $s_1$, $s_2$ and $-s_I$. As our notation suggests, the  coefficients $\gamma_{s_I}^{s_1, s_2}$ depend on the conformal spins $s_i$ (or equivalently bulk helicities) of particles $1,2,I$, but crucially not on the conformal weights $\Delta_i$. They generally further depend on color, flavor, or other internal quantum numbers, but we suppress this dependence in our notation. Helicity contributions satisfying $s_1+s_2-s_I < 0$ give rise to anti-holomorphic singularities which will be treated in subsection \ref{subsec:general-ope}.   Finally, the coupling with $s_1+s_2 = s_I$ is an edge case that we generally ignore and only arises for pure scalar $\phi^3$ interactions. Note that then barring this pure scalar interaction, we have $p_{12I} \geq 0$. 

Inserting \eqref{general-single-particle-OPE} into the generalized Wick theorem \eqref{generalizedwick-1} gives
\begin{equation} \label{single-particle-OPE-general}
\begin{aligned}
& \mc{O}_1(z_1, \bar{z}_1) : \mc{O}_2 \mc{O}_3:(z_3, \bar{z}_3) \\
& = \oint_{z_3} \frac{dz_2}{2 \pi i  z_{23}} \oint_{\bar{z}_3} \frac{d \bar{z}_2}{2 \pi i \bar{z}_{23}}  \left[\sum_{I}\frac{\gamma_{s_I}^{s_1, s_2}}{z_{12}} \sum_{m=0}^\infty  \frac{1}{m!} B(2\bar{h}_1 +  p_{12I}  + m, 2\bar{h}_2+p_{12I}) \bar{z}_{12}^{p_{12I} + m}  \partial_{\bar{z}_2}^m  \mc{O}_{I}   (z_2,\bar{z}_2)  \mc{O}_3(z_3, \bar{z}_3) \right. \\
\ &  \quad \quad  \left. +   \sum_{I}\frac{\gamma_{s_I}^{s_1, s_3}}{z_{13}} \sum_{m=0}^\infty  \frac{1}{m!} B(2\bar{h}_1 +  p_{13I}  + m, 2\bar{h}_3 +  p_{13I} ) \bar{z}_{13}^{p_{13I} + m}  \mc{O}_2 (z_2, \bar{z}_2) \partial_{\bar{z}_3}^m \mc{O}_{I} (z_3,\bar{z}_3)\right]   + \mc{O} \big( z_{13}^0 \big).
\end{aligned}
\end{equation} 
Differentiating \eqref{general-single-particle-OPE} with respect to the first anti-holomorphic coordinate gives\footnote{Note that the factorial in the denominator implies that $n \geq m-p_{12I}$.} 
\begin{equation}
\begin{aligned}
&  \LP \bar{\D}^m \mc{O}_1 \RP (z_1, \bar{z}_1) \mc{O}_2(z_2, \bar{z}_2) \\
&  \quad \quad = \sum_{I}\frac{\gamma_{s_I}^{s_1, s_2}}{z_{12}} \sum_{n=0}^\infty \frac{B(2\bar{h}_1 +p_{12I}  + n, 2\bar{h}_2 +p_{12I} )(p_{12I} + n)!}{n!(p_{12I} + n - m)!}  \bar{z}_{12}^{p_{12I} + n - m} \bar{\D}^n \mc{O}_{I} (z_2,\bar{z}_2) 
+\mc{O}\big( z_{12}^0 \big) .
\end{aligned}
\end{equation} 
Substituting this into \eqref{single-particle-OPE-general} and simplifying, we find that the $\mc{O}_1:\mc{O}_2 \mc{O}_3:$ OPE includes both single-particle and multi-particle contributions of the form
\begin{equation}\label{generalOPE-holomorphic}
\begin{split}
& \mc{O}_1(z_1, \bar{z}_1) : \mc{O}_2 \mc{O}_3:(z_3, \bar{z}_3) \\
& = \sum_{I,J} \frac{\gamma_{s_I}^{s_1, s_2} \gamma_{s_J}^{s_{I}, s_3}}{z^2_{13}} \sum_{m=0}^\infty \frac{\bar{z}_{13}^{p_{12I}+p_{I3J}+ m}}{m!}  B(2 \bar{h}_1 +p_{12I}+p_{I3J}+m,2 \bar{h}_2 +p_{12I}) \\& \quad \quad \quad \quad \quad \quad \quad \quad \quad \quad \quad \quad \quad \quad \quad \quad \quad  \times   B(   2\bar{h}_I  +p_{I3J}+m, 2\bar{h}_3  +p_{I3J}) \bar{\D}^m \mc{O}_{J} (z_3,\bar{z}_3)  \\
& \quad  + \sum_{I}\frac{\gamma_{s_I}^{s_1, s_2}}{z_{13}} \sum_{m=0}^\infty  \frac{\bar{z}_{13}^{p_{12I} + m}}{m!} B(2\bar{h}_1 + p_{12I}+ m, 2\bar{h}_2 + p_{12I}) : \LP \bar{\D}^m \mc{O}_{I} \RP  \mc{O}_3:(z_3, \bar{z}_3) \\
& \quad  + \sum_{I}\frac{\gamma_{s_I}^{s_1, s_3}}{z_{13}} \sum_{m=0}^\infty  \frac{\bar{z}_{13}^{p_{13I}+ m}}{m!} B(2\bar{h}_1 + p_{13I} +m, 2\bar{h}_3 + p_{13I}) :\mc{O}_2 (\bar{\D}^m \mc{O}_{I}):(z_3, \bar{z}_3) + \mc{O}\big( z_{13}^0 \big)\\
& \quad +\text{(terms~from~overlapping~holomorphic~\& anti-holomorphic singularities)}. 
\end{split}
\end{equation}
Note that the first sum over $m$ is a sum over anti-holomorphic descendants of the single-particle operator $\mathcal{O}_J$.  As usual, the coefficients of the descendants are fully determined by global conformal invariance in terms of the coefficient of the primary. As we will see in section \ref{sec:symmetry}, the second and third sums over $m$ differ subtly from the sum over anti-holomorphic descendants of the composites $: \mathcal{O}_I\mathcal{O}_J:$, but turn out to be a collection of multi-particle operators that transform simply under 4D bulk translations and play a key role in the analysis in section \ref{sec:symmetry}.  Retaining only the leading holomorphic singularities involving single-particle $\mathcal{O}_I$ and multi-particle $:\mathcal{O}_I \mathcal{O}_J:$ celestial operators and using ``$\sim$'' to denote this truncation, \eqref{generalOPE-holomorphic} becomes
\begin{equation}\label{generalOPE-primary-holomorphic}
\begin{split}
& \mc{O}_1(z_1, \bar{z}_1) : \mc{O}_2 \mc{O}_3:(z_3, \bar{z}_3) \\
& \sim \sum_{I,J} \gamma_{s_I}^{s_1, s_2} \gamma_{s_J}^{s_{I}, s_3}\frac{\bar{z}_{13}^{p_{12I}+p_{I3J}}}{z^2_{13}}      B(2 \bar{h}_1 +p_{12I}+p_{I3J},2 \bar{h}_2 +p_{12I})  B( 2\bar{h}_I  +p_{I3J}, 2\bar{h}_3  +p_{I3J})   \mc{O}_{J} (z_3,\bar{z}_3)  \\
& \ \ \ + \sum_{I}\gamma_{s_I}^{s_1, s_2}\frac{\bar{z}_{13}^{p_{12I}}}{z_{13}}   B(2\bar{h}_1 +p_{12I} , 2\bar{h}_2 +p_{12I}) : \mc{O}_{I} \mc{O}_3:(z_3, \bar{z}_3) \\
& \ \ \ + \sum_{I}\gamma_{s_I}^{s_1, s_3}\frac{\bar{z}_{13}^{p_{13I}}}{z_{13}}   B(2\bar{h}_1 +p_{13I} , 2\bar{h}_3 +p_{13I}) :\mc{O}_2   \mc{O}_{I}:(z_3, \bar{z}_3).
\end{split}
\end{equation} 
This expression and the method used to derive it is a major result of the paper.  Here, as above, $\gamma^{s_I, s_J}_{s_K}$ vanishes for $s_I+s_J-s_K <0$, but is otherwise proportional to the bulk coupling constant for the three-point interaction between particles of (out-going) helicity $s_I$, $s_J$ and $-s_K$.  $p_{IJK}$ is defined in \eqref{def-p} and related to the bulk scaling dimension of the associated 3-point interaction according to \eqref{p-constraint}. Note in particular this means that the operators appearing on the right-hand side carry the following conformal dimensions:
\begin{equation} \label{fusion-rule-multiparticle}
    \begin{split}
        \mathcal{O}_J ~~\text{with}~~ \Delta_J &= \Delta_1+\Delta_2+\Delta_3 + p_{12I}+p_{I3J}-2 = \Delta_1+\Delta_2+\Delta_3+s_1+s_2+s_3-s_J-4, \\
        :\mathcal{O}_I \mathcal{O}_3:~~ \text{with}~~ \Delta_I &= \Delta_1+\Delta_2+p_{12  I}-1 = \Delta_1+\Delta_2+s_1+s_2-s_I-2, \\
        :\mathcal{O}_2 \mathcal{O}_I:~~ \text{with}~~ \Delta_I &= \Delta_1+\Delta_3+p_{13  I}-1=\Delta_1+\Delta_3+s_1+s_3-s_I-2.
    \end{split}
\end{equation} 

It is interesting to note that the coefficient of the single-particle operator simplifies to the following ratio of gamma functions:
\begin{equation}
    \begin{split}
           B(2 \bar{h}_1  +p_{12I}+p_{I3J}, 2 \bar{h}_2 +p_{12I}) & B( 2\bar{h}_I  +p_{I3J}, 2\bar{h}_3  +p_{I3J}  ) \\
            &= \frac{\Gamma(2 \bar{h}_1 +p_{12I}+p_{I3J}) \Gamma(2 \bar{h}_2 +p_{12I})\Gamma( 2\bar{h}_3  +p_{I3J}) }{\Gamma( 2\bar{h}_1+2\bar{h}_2+2\bar{h}_3+  2p_{12I}  +2p_{I3J})}  \\
           & = B(2 \bar{h}_1 +p_{12I}+p_{I3J},2 \bar{h}_2 +p_{12I},2\bar{h}_3  +p_{I3J}),
    \end{split}
\end{equation}
where here $B(x,y,z)$ is the generalized Euler beta function
\begin{equation}\label{trivariatebetafunction}
    \begin{split}
        B(x,y,z) = \frac{\Gamma(x)\Gamma(y) \Gamma(z)}{\Gamma(x+y+z)} = B(x,y) B(x+y,z),
    \end{split}
\end{equation}
and we have used that $\bar{h}_I = \bar{h}_1 +  \bar{h}_2 +p_{12I}$.  The idea that multi-collinear celestial OPE coefficients take the form of generalized Euler beta functions was first put forth by \cite{Ebert:2020nqf} in the context of Yang-Mills theory and here we see that it extends to general massless interactions.  Moreover, we observe that the OPE coefficients of multi-particle OPEs inherit interesting pole structure in boost-weight space from the single-particle OPE coefficients. We comment on the relation of this pole structure to known holographic symmetry algebras in the discussion section \ref{sec:discussion}.

To compare with results for Yang-Mills theory in the literature, we set $d_V = 4$ and $s_i = \pm 1$ and we recover and extend all of the multi-particle gluon OPEs derived in \cite{Guevara:2024ixn} from bulk holomorphic collinear limits. Strictly speaking, many of the calculations in \cite{Guevara:2024ixn} determine the purely anti-holomorphic (as opposed to holomorphic) singular terms.  In these cases, our results reproduce theirs upon the purely cosmetic exchanges of $z \leftrightarrow\bar{z}$, $h\leftrightarrow \bar{h}$ and $s_i =\pm1 \leftrightarrow s_i =\mp1$.  Our OPE coefficients for single-particle contributions to the multi-particle OPE precisely match the ones in \cite{Guevara:2024ixn}.  In \cite{Guevara:2024ixn}, the OPE coefficients of multi-particle contributions were calculated up to an undetermined function $\tilde f\left(t; \frac{\omega_1}{\omega_P},\frac{\omega_2}{\omega_P},\frac{\omega_3}{\omega_P}\right)$ where  $\tilde f$ is required to satisfy 
\begin{equation}
    \begin{split}
        \int_0^1 dt~\tilde f \left(t; \frac{\omega_1}{\omega_P},\frac{\omega_2}{\omega_P},\frac{\omega_3}{\omega_P}\right)=1.
    \end{split}
\end{equation}
Allowing $\tilde f$ to have non-trivial color structure, our results are consistent with those in \cite{Guevara:2024ixn} with 
\begin{equation}
    \begin{split}
        \tilde f^{a_1a_2a_3}{}_{a_q a_\ell}\left(t; \frac{\omega_1}{\omega_P},\frac{\omega_2}{\omega_P},\frac{\omega_3}{\omega_P}\right)
        = f^{a_1a_2}{}_{a_q} \delta_{a_\ell}^{a_3} \delta\left(t- \frac{\omega_1+\omega_2}{\omega_P}\right) + f^{a_1a_3}{}_{a_\ell} \delta_{a_q}^{a_2} \delta\left(t- \frac{\omega_2}{\omega_P}\right).
    \end{split}
\end{equation}
Note both color structures appearing on the right-hand side were found in \cite{Guevara:2024ixn} and our results match provided that a different (color-independent) $\tilde f$ is chosen for each term.  This minor modification appears to be consistent with the overall analysis in \cite{Guevara:2024ixn}.  

Similarly, our earlier result \eqref{GGGplusOPE-final} is a special case of \eqref{generalOPE-primary-holomorphic} obtained by taking $d_V =5$, $s_i=2$, and $\LP h_i, \bar{h}_i \RP = \LP \frac{\Delta_i}{2} + 1, \frac{\Delta_i}{2} - 1 \RP$ as in \eqref{G^+G^+-primary} and further setting $\kappa = - 2 \gamma_2^{2,2}$.  Finally, for ease of comparison with bulk collinear limits in section \ref{sec:bulk-collinear-limits}, we obtain the holomorphic singular terms of celestial OPEs in Einstein gravity from \eqref{generalOPE-primary-holomorphic} by setting $s_i = \pm 2$ and taking the only non-vanishing coupling constants to be $\gamma_{\pm2 }^{2,\pm2} =\gamma^{-2, 2}_{-2} =  - \kappa/2$
\begin{equation} \label{boundary-graviton-holomorphic}
    \begin{split}
       G_{\Delta_1}^{s_1}& (z_1,  \bar{z}_1) :G_{\Delta_2}^{s_2} G_{\Delta_3}^{s_3}:(z_3, \bar{z}_3)\\   
      & \sim \frac{\kappa^2}{4} \frac{\bar{z}_{13}^2}{z_{13}^2}\delta_{s_1 + s_2 +s_3 \geq 0} 
      \sum_{m=0}^\infty \frac{\bar{z}_{13}^m}{m!}B(2 \bar{h}_1 + 2+m, 2 \bar{h}_2 + 1, 2 \bar{h}_3 + 1) \bar{\partial}^mG_{\Delta_1 + \Delta_2 + \Delta_3}^{ \min( s_1, s_2, s_3)  }(z_3, \bar{z}_3) \\
      \ & \ \ \ - \frac{\kappa}{2} \frac{\bar{z}_{13}}{z_{13}} \delta_{s_1 + s_2 \geq 0} \sum_{m=0}^\infty \frac{\bar{z}_{13}^m}{m!} B(2 \bar{h}_1 + 1+m, 2 \bar{h}_2 + 1) :\Big(\bar{\partial}^m G^{ \min(s_1, s_2) } _{\Delta_1 + \Delta_2} \Big) G_{\Delta_3}^{s_3}: (z_3, \bar{z}_3)  \\
\ & \ \ \ - \frac{\kappa}{2} \frac{\bar{z}_{13}}{z_{13}}   \delta_{s_1 + s_3 \geq 0}\sum_{m=0}^\infty \frac{\bar{z}_{13}^m}{m!}B(2 \bar{h}_1 + 1+m, 2 \bar{h}_3 + 1) :G_{\Delta_2}^{s_2} \Big(\bar{\partial}^m G^{ \min(s_1, s_3) }_{\Delta_1 + \Delta_3}\Big):(z_3, \bar{z}_3).
    \end{split}
\end{equation} 

\subsection{All singularities for general helicity}\label{subsec:general-ope}

In the previous subsection, we neglected contributions arising from anti-holomorphic singularities in the single-particle OPEs \eqref{general-single-particle-OPE}; this is equivalent to the usual bulk prescription of calculating collinear amplitudes from holomorphic multi-collinear limits. Anti-holomorphic singularities can only arise in a single-particle OPE $\mathcal{O}_1\mathcal{O}_2 \to \mathcal{O}_I$ if $s_1+s_2\leq s_I$.  Note that for asymptotic massless particles with helicities $|s_i| <2$, this condition can never be satisfied by the $G^+G^+$ OPE \eqref{G^+G^+-primary}, so our result in subsection \ref{subsec:positive-helicity-gravitons} is uncorrected. 

However, in general these anti-holomorphic poles contribute to the contour integrals in the generalized Wick theorem \eqref{generalizedwick-1}. The contribution to the single-particle OPE \eqref{general-single-particle-OPE} can be readily deduced by simply augmenting it by terms that exchange $z\leftrightarrow\bar{z}$ and $h \leftrightarrow \bar{h}$  and replacing $p_{IJK} \to \bar{p}_{IJK}\equiv -s_I-s_J+s_K-1$: 
 \begin{equation}\label{general-single-particle-OPE-conjugation-symmetric}
\begin{aligned}
\mc{O}_1(z_1, \bar{z}_1) \mc{O}_2(z_2, \bar{z}_2) & = \frac{1}{z_{12}} \sum_I \sum_{n=0}^\infty  \frac{\gamma_{s_I}^{s_1, s_2}}{n!} B(2 \bar{h}_1+ p_{12I}+ n, 2 \bar{h}_2+ p_{12I}) \bar{z}_{12}^{p_{12I} + n} \bar{\D}^n \mc{O}_I (z_2,\bar{z}_2) \\
& \ \ \ + \frac{1}{\bar{z}_{12}} \sum_I \sum_{n=0}^\infty  \frac{\bar{\gamma}_{s_I}^{s_1, s_2}}{n!} B( 2h_1+\bar{p}_{12I}+ n, 2h_2+\bar{p}_{12I}) z_{12}^{\bar{p}_{12I} + n} \D^n \mc{O}_I (z_2,\bar{z}_2) \\
\ & \ \ \ + \mc{O} \LP z_{12}^0 \bar{z}_{12}^0 \RP,
\end{aligned}
\end{equation}
where $\bar{\gamma}_{s_I}^{s_1, s_2}$ vanishes for $s_1+s_2-s_I>0$, again we generally ignore the edge case $s_1+s_2-s_I=0$, and $\bar{\gamma}_{s_I}^{s_1, s_2} = \gamma_{-s_I}^{-s_1, - s_2}$ upon appropriate conjugation of all suppressed quantum numbers.   Here $\bar{p}_{12I}$, like $p_{12I}$, is related to the bulk scaling dimension $d_V$ of the associated 3-point interaction by $\bar{p}_{12I}= d_V-4$.

The result of applying the generalized Wick theorem to the full single-particle OPE \eqref{general-single-particle-OPE-conjugation-symmetric} is \emph{almost} given by simply augmenting the result \eqref{generalOPE-primary-holomorphic} for holomorphic singularities by terms that exchange $z\leftrightarrow\bar{z}$ and $h \leftrightarrow \bar{h}$ and replace $p_{IJK} \to \bar{p}_{IJK}$.  The only subtlety is that the generalized Wick formula is sensitive to overlapping holomorphic and anti-holomorphic singularities that arise between the two OPEs in the procedure and were therefore missed by the analysis in subsection \ref{subsec:holomorphic-allspin}.

To illustrate this subtlety, consider the following explicit example in Einstein-Yang-Mills.  Beginning with the single-particle OPEs \cite{Fan:2019emx,Pate:2019lpp}
\begin{equation}
    \begin{split}
        G^+_{\Delta_1}(z_1, \bar{z}_1) O^{\pm a}_{\Delta_2}(z_2, \bar{z}_2) 
            &\sim -\frac{\kappa}{2} \frac{\bar{z}_{12}}{z_{12}}B(\Delta_1-1, \Delta_2\mp 1+1) O^{\pm a}_{\Delta_1+\Delta_2}(z_2, \bar{z}_2) , \\
        O^{+a}_{\Delta_1}(z_1, \bar{z}_1) O^{-b}_{\Delta_2}(z_2, \bar{z}_2)
            & \sim - \frac{if^{ab}{}_c}{z_{12}} B(\Delta_1-1, \Delta_2+1) O^{-c}_{\Delta_1+\Delta_2-1}(z_2, \bar{z}_2) \\& \quad - \frac{if^{ab}{}_c}{\bar{z}_{12}} B(\Delta_1+1, \Delta_2-1)O^{+c}_{\Delta_1+\Delta_2-1}(z_2, \bar{z}_2),
    \end{split}
\end{equation}
then to compute the multi-particle OPE $G^+  : O^{+a} O^{-b}:$, we need in particular
\begin{equation}
    \begin{split}
        \oint_{z_3} &\frac{dz_{2}}{2 \pi i } \oint_{\bar{z}_3} \frac{d \bar{z}_2}{2 \pi i } \frac{1}{z_{23}} \frac{1}{\bar{z}_{23}} \wick{\c G^+_{\Delta_1}(z_1, \bar{z}_1) \c O^{+a}_{\Delta_2}(z_2, \bar{z}_2} )O^{-b}_{\Delta_3}(z_3, \bar{z}_3)\\
        &\supset-\frac{\kappa}{2}\oint_{z_3}\frac{dz_{2}}{2 \pi i }  \oint_{\bar{z}_3} \frac{d \bar{z}_2}{2 \pi i } \frac{1}{z_{23}} \frac{1}{\bar{z}_{23}}
         \frac{\bar{z}_{12}}{z_{12}}\sum_{m=0}^{\infty} \frac{1}{m!}B(\Delta_1-1+m, \Delta_2 ) \bar{z}_{12}^m \partial_{\bar{z}_2}^m \\& \quad\quad\quad\quad\quad\quad\quad\quad \times  \left[  - \frac{if^{ab}{}_c}{\bar{z}_{23}} B( \Delta_1+\Delta_2+1,\Delta_3-1)O^{+c}_{\Delta_1+\Delta_2+\Delta_3-1}(z_3, \bar{z}_3)\right]\\
        &= -\frac{\kappa}{2}  \frac{if^{ab}{}_c}{z_{13}} B(\Delta_1-1,\Delta_2-1) B( \Delta_1+\Delta_2+1,\Delta_3-1)O^{+c}_{\Delta_1+\Delta_2+\Delta_3-1}(z_3, \bar{z}_3) .
    \end{split}
\end{equation}
As demonstrated above, the overlapping holomorphic and anti-holomorphic singularities in  the generalized Wick theorem give rise to simple (as opposed to double) poles in the multi-particle OPE with single-particle operator coefficients. 
 
More generally, these overlapping poles give non-vanishing contributions when an OPE channel $1 2 \to I$ with an anti-holomorphic (holomorphic) pole with  $\bar{p}_{12  I} \geq 1$  ($p_{12I}\geq 1$) composes with an OPE channel $I3\to J$ with a holomorphic (anti-holomorphic) pole with $ p_{I3J}=0$ ($\bar{p}_{I3J}=0$), where $p_{IJK}$ is defined as in \eqref{def-p} and related to the bulk scaling dimension of the corresponding three-point interaction by $d_V = p_{IJK}+4$. In particular, there are no contributions of this type in pure tree-level Yang-Mills ($d_V = 4$) or Einstein gravity ($d_V = 5$) , but as seen above such contributions do arise in Einstein-Yang-Mills. 

Our general solution that accounts for all possible holomorphic and anti-holomorphic singularities is 
\begin{equation}\label{generalOPE-primary-holomorphic-and-antiholomorphic}
\begin{split}
 \mc{O}_1&(z_1, \bar{z}_1) : \mc{O}_2 \mc{O}_3:(z_3, \bar{z}_3) \\
& \sim \sum_{I,J} \frac{\gamma_{s_I}^{s_1, s_2} \gamma_{s_J}^{s_{I}, s_3}}{z^2_{13}}    \bar{z}_{13}^{p_{12I}+p_{I3J}}  B(2 \bar{h}_1 + p_{12I} + p_{I3J}, 2\bar{h}_2 + p_{12I}, 2\bar{h}_3 + p_{I3J})   \mc{O}_{J} (z_3,\bar{z}_3)  \\
& \quad+ \sum_{I,J} \frac{\bar{\gamma}_{s_I}^{s_1, s_2} \bar{\gamma}_{s_J}^{s_{I}, s_3}}{\bar{z}^2_{13}}    z_{13}^{\bar{p}_{12I}+\bar{p}_{I3J}}  B(2 h_1 +\bar{p}_{12I}+\bar{p}_{I3J}, 2h_2 +\bar{p}_{12I}, 2h_3+\bar{p}_{I3J})   \mc{O}_{J} (z_3,\bar{z}_3) \\
& \quad  +\sum_{I,J}\delta_{\bar{p}_{I3J},0}
\frac{\gamma^{s_1,s_2}_{s_I}  \bar{\gamma}_{s_J}^{s_I,s_3}}{z_{13}}\bar{z}_{13}^{p_{12I}-1}B(2h_1+ 2h_2 -2,2h_3)
 \\& \quad \quad\quad \quad\quad \quad\quad \quad\quad \quad \times \left[B  (2\bar{h}_1+p_{12I}-1,2\bar{h}_2+p_{12I}) - B (2\bar{h}_1+p_{12I}-1,2 \bar{h}_2 )  \right]{\mathcal{O}}_{J}(z_3, \bar{z}_3)
           \\
& \quad  +\sum_{I,J}\delta_{p_{I3J},0}
\frac{\bar{\gamma}^{s_1,s_2}_{s_I}   \gamma_{s_J}^{s_I,s_3}}{\bar{z}_{13}}z_{13}^{\bar{p}_{12I}-1}B(2\bar{h}_1+ 2\bar{h}_2 -2,2\bar{h}_3)
 \\& \quad \quad\quad \quad\quad \quad\quad \quad\quad \quad \times \left[B  (2h_1+\bar{p}_{12I}-1,2h_2+\bar{p}_{12I}) - B (2h_1+\bar{p}_{12I}-1,2 h_2 )  \right]{\mathcal{O}}_{J}(z_3, \bar{z}_3)
           \\
&\quad+ \sum_{I}\frac{\gamma_{s_I}^{s_1, s_2}}{z_{13}}  \bar{z}_{13}^{p_{12I}} B(2\bar{h}_1 +p_{12I}, 2\bar{h}_2 +p_{12I}) : \mc{O}_{I} \mc{O}_3:(z_3, \bar{z}_3) \\
& \quad+ \sum_{I}\frac{\bar{\gamma}_{s_I}^{s_1, s_2}}{\bar{z}_{13}}  z_{13}^{\bar{p}_{12I}} B(2h_1 +\bar{p}_{12I}, 2h_2 +\bar{p}_{12I}) : \mc{O}_{I} \mc{O}_3:(z_3, \bar{z}_3) \\
&\quad + \sum_{I}\frac{\gamma_{s_I}^{s_1, s_3}}{z_{13}}  \bar{z}_{13}^{p_{13I}} B(2\bar{h}_1 +p_{13I}, 2\bar{h}_3 +p_{13I}) :\mc{O}_2   \mc{O}_{I}:(z_3, \bar{z}_3) \\
& \quad + \sum_{I}\frac{\bar{\gamma}_{s_I}^{s_1, s_3}}{\bar{z}_{13}}  z_{13}^{\bar{p}_{13I}} B(2h_1 +\bar{p}_{13I}, 2h_3+\bar{p}_{13I}) :\mc{O}_2   \mc{O}_{I}:(z_3, \bar{z}_3),
\end{split}
\end{equation}
where here we have only kept the singular contributions from single-particle celestial primaries $\mathcal{O}_I$ and multi-particle operators $:\mathcal{O}_I\mathcal{O}_J:$.

\section{Multi-particle graviton OPEs from bulk collinear limits}
\label{sec:bulk-collinear-limits}

In this section, we reproduce the boundary calculations of the previous section from bulk collinear limits in Einstein gravity at tree-level. More specifically, using the methods developed in \cite{Ball:2023sdz} and \cite{Guevara:2024ixn}, we determine the contribution from single-graviton operators to the multi-graviton OPE in Einstein gravity in subsection \ref{subsec:bulk-single-particle}. The contributions from composite graviton operators are more subtle so in subsection \ref{subsec:bulk-multi-particle} we use a known recursion relation for MHV graviton amplitudes to determine the composite contribution in the MHV sector. 

In each instance, we obtain exact results from bulk scattering that precisely agree with the results in the previous section.  We do not perform the most general bulk calculation that fully reproduces every case of our boundary formula \eqref{generalOPE-primary-holomorphic}, but the examples we study incorporate several non-trivial features that do not appear in the Yang-Mills analysis in \cite{Guevara:2024ixn}, whose results we already reproduce.  Specifically, these features include the somewhat exotic $\bar{z}^p/z$ with $p>0$ singularity structure exhibited by many single-particle celestial OPEs, the more complex singularity structure of mixed-helicity holomorphic collinear limits and the additional technical challenges arising from treating MHV graviton as opposed to MHV gluon amplitudes. 

\subsection{Single-graviton contributions to the multi-particle graviton OPE}
\label{subsec:bulk-single-particle} 

The authors of \cite{Ball:2023sdz} define a notion of a multi-holomorphic collinear limit, \textit{i.e.} a holomorphic collinear limit in which multiple momenta are taken asymptotically parallel in some prescribed ordering. For our application to multi-particle OPEs, we are generally interested in the singular behavior of a scattering amplitude in the limit $|\hat{p}_2 - \hat{p}_3| \ll |\hat{p}_1 - \hat{p}_3| \ll 1$, which, after removing the singular part of the $2 \to 3$ limit, is dual to the boundary OPE $\mathcal{O}_1(z_1, \bar{z}_1) : \mathcal{O}_2 \mathcal{O}_3:(z_3, \bar{z}_3)$.  Bulk collinear limits require further specification and here we follow the prescription in \cite{Ball:2023sdz}, in which holomorphic spinor-helicity variables are taken parallel, corresponding to holomorphic poles on the boundary. Using the conventions for spinor helicity variables detailed in appendix \ref{sec:conventions}, we parametrize the holomorphic Weyl spinors of the to-be-collinear outgoing external states as
\begin{equation}
\begin{aligned}
\KET{1} & = \sqrt{2 \omega_1} \LP \KET{\hat{3}} + \epsilon \KET{r} \RP, \\
\KET{2} & = \sqrt{2 \omega_2} \LP \KET{\hat{3}} + \eta \epsilon \KET{r} \RP, \\
\KET{3} & = \sqrt{2 \omega_3} \KET{\hat{3}},
\end{aligned}
\end{equation}
where $\KET{r}$ is the reference spinor \begin{equation}
\KET{r} = \begin{bmatrix} 0 \\ 1 \end{bmatrix}
\end{equation}
and 
\begin{equation}
 \KET{\hat{3}} = \begin{bmatrix} -1 \\ -z_3  \end{bmatrix}.
\end{equation}
Equivalently in terms of holomorphic coordinates
\begin{equation}
\begin{aligned}
z_1 & = z_3 - \epsilon, \\
z_2 & = z_3 - \eta \epsilon.
\end{aligned}
\end{equation}
The multi-collinear limit $\hat{p}_1 \approx \hat{p}_2 \approx \hat{p}_3$ corresponds to taking $\epsilon \to 0$,\footnote{Note that the parameter $\epsilon$ is unrelated to the Levi-Civita symbol $\varepsilon$ used to raise and lower spinor indices.} while the value of $\eta$ determines the order of the multi-collinear limit, \textit{i.e.}, which two of the three momenta are parametrically closer to one another than they are to the third. Concretely, we have \begin{equation}
\begin{cases}
\eta \to 0 & \implies |\hat{p}_2 - \hat{p}_3| \ll |\hat{p}_1 - \hat{p}_3| \approx |\hat{p}_1 - \hat{p}_2|, \\
\eta \to 1 & \implies |\hat{p}_1 - \hat{p}_2| \ll |\hat{p}_1 - \hat{p}_3| \approx |\hat{p}_2 - \hat{p}_3|, \\
\eta \to \infty & \implies |\hat{p}_1 - \hat{p}_3| \ll |\hat{p}_1 - \hat{p}_2| \approx |\hat{p}_2 - \hat{p}_3|.
\end{cases}
\end{equation}
Mechanically, this convenient parametrization allows one to extract the leading-order multi-collinear divergence of the scattering amplitude by computing the leading $\mathcal{O}\LP \epsilon^{-2} \RP$ term in the limit $\epsilon \to 0$, and only subsequently taking an appropriate limit of $\eta$. In the limit $\epsilon \to 0$, the tree-level $n$-particle amplitude factorizes and the splitting function $\mbox{Split}[1^{s_1} 2^{s_2} 3^{s_3} \to J^{s_J} ]$ is then defined to be the $\epsilon^{-2}$ coefficient of the $(n-2)$-particle amplitude. 

Following \cite{Ball:2023sdz}, this leading-order divergence of the unstripped amplitude $\mc{A}_n$ can be written as a sum over factorization channels
\begin{equation}
\begin{split}
  \mathcal{A}_n \LP 1^{s_1} 2^{s_2} 3^{s_3} \cdots n^{s_n} \RP  
& = \sum_J \mbox{Split}[1^{s_1} 2^{s_2} 3^{s_3} \to J^{s_J} ] \mathcal{A}_{n-2} \LP J^{s_J} \ldots n^{s_n} \RP + \mathcal{O}\LP \epsilon^{-1} \RP \\
\ & = D_{1,2} + D_{2,3} + D_{1,3} + \mathcal{O}\LP \epsilon^{-1} \RP.
\end{split}
\end{equation}
Here $D_{i,j}$ is the sum over all diagrams that factorize on the propagators $s_{ij}$ and $s_{123}$, where $s_{i_1 \cdots i_k}$ is the generalized Mandelstam invariant $s_{i_1 \cdots i_k} = - (\sum_{n=1}^k p_{i_n})^2$, not to be confused with 2D conformal spin $s_i$.  Also, here and throughout, $\mathcal{A}_n$ is used to denote the unstripped amplitude, meaning the amplitude including the delta-function for momentum conservation. Mellin-transforming with respect to the energy of each external particle then yields an expression for the leading singularities in the corresponding boundary OPE with contributions from all possible channels. For example, the authors of \cite{Ball:2023sdz} find that the all-plus graviton splitting function is  
\begin{equation}\label{++++splitting}
\begin{aligned}
\mbox{Split}[1^{++} 2^{++} 3^{++} \to J^{++}] & = \frac{\kappa^2}{4 \epsilon^2} \frac{(\omega_1 + \omega_2 + \omega_3)^2 \LP \bar{z}_{12} \frac{(\omega_1 \bar{z}_{13} + \omega_2 \bar{z}_{23})^2}{\omega_1 \omega_2 (1 - \eta)} + \bar{z}_{23} \frac{(\omega_2 \bar{z}_{12} + \omega_3 \bar{z}_{13})^2}{\omega_2 \omega_3 \eta} + \bar{z}_{13} \frac{(\omega_1 \bar{z}_{12} - \omega_3 \bar{z}_{23})^2}{\omega_1 \omega_3}
\RP}{\omega_1 \omega_2 (1 - \eta) \bar{z}_{12} + \omega_1 \omega_3 \bar{z}_{13} + \omega_2 \omega_3 \eta \bar{z}_{23}} \\
\ & = \frac{\kappa^2}{4 \epsilon^2} \frac{(\omega_1 + \omega_2 + \omega_3)^2}{\omega_1 \omega_2 \omega_3} \LP \omega_1 \frac{\bar{z}_{12} \bar{z}_{13}}{1 - \eta} + \omega_2 \frac{\bar{z}_{12} \bar{z}_{23}}{\eta(1 - \eta)} + \omega_3 \frac{\bar{z}_{13} \bar{z}_{23}}{\eta} \RP.
\end{aligned}
\end{equation}
To perform the desired Mellin transform, we make the change of variables  
\begin{equation}\label{cov}
\begin{aligned}
\omega & \equiv \omega_1 + \omega_2 + \omega_3, \\
\sigma_1 & \equiv \omega_1 / \omega, \\
\sigma_2 & \equiv \omega_2 / \omega, \\
\end{aligned}
\end{equation}
and define the homogenized splitting function \begin{equation}\label{splithat}
\widehat{\mbox{Split}}[1^{s_1} 2^{s_2} 3^{s_3} \to J^{s_J}] \equiv \omega^{-(s_1 + s_2 + s_3 - s_J - 4)} \mbox{Split}[1^{s_1} 2^{s_2} 3^{s_3} \to J^{s_J}].
\end{equation}
Then, the Mellin-transformed amplitude
\begin{equation}
    \begin{split}
        \widetilde{\mathcal{A}}_n& \LP \Delta_1, s_1, z_1, \bar{z}_1; \Delta_2, s_2, z_2, \bar{z}_2; \Delta_3, s_3, z_3, \bar{z}_3; \ldots \RP 
       \equiv \left(\prod_{i = 1}^n \int_0^\infty \frac{d \omega_i}{\omega_i} \omega_i^{\Delta_i}\right)\mathcal{A}_n \LP 1^{s_1} 2^{s_2} 3^{s_3} \cdots n^{s_n} \RP 
    \end{split}
\end{equation}
is given by
\begin{equation}
    \begin{split}
       \widetilde{\mathcal{A}}_n& \LP \Delta_1, s_1, z_1, \bar{z}_1; \Delta_2, s_2, z_2, \bar{z}_2; \Delta_3, s_3, z_3, \bar{z}_3; \ldots \RP \\ 
        & =  \int_0^1 d\sigma_1 \int_0^{1-\sigma_1} d\sigma_2 ~\sigma_1^{\Delta_1 - 1} \sigma_2^{\Delta_2 - 1} (1 {-} \sigma_1 {-} \sigma_2)^{\Delta_3 - 1}\sum_J\widehat{\mbox{Split}}[1^{s_1} 2^{s_2} 3^{s_3} \to J^{s_J}]\\& \quad  \quad \times \sum_{m=0}^\infty \frac{(\sigma_1 \bar{z}_{13} + \sigma_2 \bar{z}_{23})^m}{m!} \partial_{\bar{z}_3}^m  \widetilde{\mathcal{A}}_{n-2} \LP \Delta_1 {+} \Delta_2 {+} \Delta_3 {+} s_1 {+} s_2 {+} s_3 {-} s_J {-} 4, s_J, z_3, \bar{z}_3; \ldots \RP  + \mc{O}\LP \epsilon^{-1} \RP.
    \end{split}
\end{equation}
Note that the conformal weight for the first particle in $\widetilde{\mathcal{A}}_{n-2}$ is consistent with the fusion rule for the single-particle contribution to the multi-particle OPE derived in the previous section \eqref{fusion-rule-multiparticle}.  

According to the celestial holographic dictionary, $\widetilde{\mathcal{A}}_n$ is interpreted as an $n$-point correlation function of boundary celestial operators, and the $\epsilon^{-2}$ singularity encodes single-particle contributions to the multi-particle celestial OPE. For the case of all-positive-helicity gravitons, we find in particular 
\begin{equation}
\begin{aligned}
G_{\Delta_1}^+&(z_1, \bar{z}_1) G_{\Delta_2}^+(z_2, \bar{z}_2) G_{\Delta_3}^+(z_3, \bar{z}_3)\\
& = \frac{\kappa^2}{4} \sum_{m=0}^\infty \sum_{\ell=0}^m \frac{\bar{z}_{13}^\ell \bar{z}_{23}^{m-\ell}}{\ell! (m-\ell)!} \bar{\partial}^m G^+_{\Delta_1 + \Delta_2 + \Delta_3} (z_3, \bar{z}_3)   \Bigg[ B(\Delta_1 + \ell, \Delta_2 - 1 + m - \ell, \Delta_3 -1) \frac{\bar{z}_{12} \bar{z}_{13}}{z_{12} z_{13}} \\
\ & \ \ \ + B(\Delta_1 - 1 + \ell, \Delta_2 + m- \ell, \Delta_3 - 1) \frac{\bar{z}_{12} \bar{z}_{23}}{z_{12} z_{23}}  + B(\Delta_1 - 1 + \ell, \Delta_2 - 1 + m - \ell, \Delta_3) \frac{\bar{z}_{13} \bar{z}_{23}}{z_{13} z_{23}} \Bigg] \\
\ & \ \ \ + \mbox{(multi-particle contributions)}. 
\end{aligned}
\end{equation}
To extract $G_{\Delta_1}^+ :G_{\Delta_2}^+ G_{\Delta_3}^+:$ from this expression, we decompose $(z_{12}, \bar{z}_{12}) = (z_{13} - z_{23}, \bar{z}_{13} - \bar{z}_{23})$, treat the whole expression formally as a function of four independent variables $ \LP z_{13}, \bar{z}_{13}, z_{23}, \bar{z}_{23} \RP$, then expand in the latter two and extract the term of order $\Theta \LP z_{23}^0 \bar{z}_{23}^0 \RP$, where $\Theta$ is defined in the text above \eqref{OPE-decomp-explained}. Noting that the leading rational term $\bar{z}_{13}^\ell \bar{z}_{23}^{m-\ell}$ for $m \geq \ell$ is regular in $\bar{z}_{23}$, and that the second and third terms in the brackets vanish as $\bar{z}_{23} \to 0$, we find that terms of order $\Theta \LP z_{23}^0 \bar{z}_{23}^0 \RP$ only receive contributions from the first line in brackets. Expanding this term and throwing away all but the order  $z_{23}^0\bar{z}_{23}^0$ contribution then gives \begin{equation}
\begin{aligned}
G_{\Delta_1}^+(z_1, \overline{z}_1) :G_{\Delta_2}^+ G_{\Delta_3}^+: (z_3, \overline{z}_3) & = \frac{1}{z_{13}^2} \frac{\kappa^2}{4} \sum_{m=0}^\infty \frac{\overline{z}_{13}^{m+2}}{m!}B(\Delta_1 + m, \Delta_2 - 1, \Delta_3 -1)  \bar{\partial}^m  G^+_{\Delta_1 + \Delta_2 + \Delta_3} (z_3, \overline{z}_3)  \\
\ & \quad + \mbox{(multi-particle contributions)}.
\end{aligned}
\end{equation}
After application of the generalized beta function identity \eqref{trivariatebetafunction}, this correctly reproduces the single-particle term in \eqref{GGGplusOPE-final}.

This analysis readily generalizes to all possible helicity combinations of gravitons, where we need only substitute the appropriate splitting function. For example, straightforward diagrammatic analysis of the leading factorization channels $D_{i,j}$ gives 
\begin{equation}
    \mbox{Split}[1^{++} 2^{++} 3^{--} \to J^{++}]= 0,
\end{equation}
while more interestingly \begin{equation}
\begin{aligned} 
\mbox{Split}[1^{++} 2^{++} 3^{--} \to J^{--}]
& = \frac{\kappa^2 \omega_3^3}{4 \epsilon^2 \eta (1-\eta)} \times \frac{\eta \omega_1 \bar{z}_{12} \bar{z}_{13} + \omega_2 \bar{z}_{12} \bar{z}_{23} + (1 - \eta) \omega_3 \bar{z}_{13} \bar{z}_{23}}{\omega_1 \omega_2 (\omega_1 + \omega_2 + \omega_3)^2}.
\end{aligned}
\end{equation}
Under a Mellin transformation, this yields
\begin{equation}
\begin{aligned}
  G_{\Delta_1}^+&(z_1, \bar{z}_1) G_{\Delta_2}^+(z_2, \bar{z}_2) G_{\Delta_3}^-(z_3, \bar{z}_3) \\
& = \frac{\kappa^2}{4} \sum_{m=0}^\infty \sum_{\ell=0}^m \frac{\bar{z}_{13}^\ell \bar{z}_{23}^{m-\ell}}{\ell!(m-\ell)!} \bar{\partial}^m G^-_{\Delta_1 + \Delta_2 + \Delta_3}(z_3, \bar{z}_3)   \left[ B(\Delta_1 + \ell, \Delta_2 - 1 + m - \ell, \Delta_3 + 3) \frac{\bar{z}_{12} \bar{z}_{13}}{z_{12}z_{13} }\right. \\
\ & \ \ \left.+   B(\Delta_1 - 1 + \ell, \Delta_2 + m - \ell, \Delta_3 + 3) \frac{\bar{z}_{12} \bar{z}_{23}}{z_{12} z_{23}}  +  B(\Delta_1 - 1 + \ell, \Delta_2 - 1 + m - \ell, \Delta_3 + 4) \frac{\bar{z}_{13} \bar{z}_{23}}{z_{13} z_{23}} \right]  \\
\ & \ \ + (\mbox{multi-particle contributions}).
\end{aligned}
\end{equation}
Generalizing to any set of graviton helicities and assembling the result into a single expression gives  
\begin{equation}
\begin{aligned}
 G_{\Delta_1}^{s_1}&(z_1, \bar{z}_1) G_{\Delta_2}^{s_2}(z_2, \bar{z}_2) G_{\Delta_3}^{s_3}(z_3, \bar{z}_3) \\
& = \delta_{s_1 {+} s_2 {+} s_3 \geq 0} \frac{\kappa^2}{4} \sum_{m=0}^\infty \sum_{\ell=0}^m \frac{\bar{z}_{13}^\ell \bar{z}_{23}^{m-\ell}}{\ell!(m{-}\ell)!} \bar{\partial}^m G^{ \min(s_1, s_2, s_3) }_{\Delta_1 {+} \Delta_2 {+} \Delta_3}(z_3, \bar{z}_3) \\& \quad 
\times \left[ B(2 \bar{h}_1 {+} 2 {+} \ell, 2 \bar{h}_2 {+} 1 {+} m {-} \ell, 2 \bar{h}_3 {+} 1) \frac{\bar{z}_{12} \bar{z}_{13}}{z_{12}z_{13} } +   B(2 \bar{h}_1 {+} 1 {+} \ell, 2 \bar{h}_2 {+} 2 {+} m {-} \ell, 2 \bar{h}_3 {+} 1) \frac{\bar{z}_{12} \bar{z}_{23}}{z_{12} z_{23}} \right. \\& \quad 
\left. \quad \quad  
+  B(2 \bar{h}_1 {+} 1 {+} \ell, 2 \bar{h}_2 {+} 1 {+} m{-} \ell, 2 \bar{h}_3 {+} 2) \frac{\bar{z}_{13} \bar{z}_{23}}{z_{13} z_{23}} \right]    + (\mbox{multi-particle contributions}).
\end{aligned}
\end{equation}
Then, removing divergences in $z_{23}$ and further expanding in $z_{13}$ and $\bar{z}_{13}$, we find 
\begin{equation}\begin{split}
& G_{\Delta_1}^{s_1}(z_1,  \bar{z}_1) :G_{\Delta_2}^{s_2}  G_{\Delta_3}^{s_3}:(z_3, \bar{z}_3) \\
& \quad = \delta_{s_1 + s_2 + s_3 \geq 0} \frac{\kappa^2}{4}  \sum_{m=0}^\infty \frac{1}{m!} \frac{\bar{z}_{13}^{2+m}}{z_{13}^2}B(2\bar{h}_1 + 2 + m, 2\bar{h}_2 + 1, 2\bar{h}_3 + 1) \bar{\partial}^m G^{ \min(s_1, s_2, s_3) }_{\Delta_1 + \Delta_2 + \Delta_3}(z_3, \bar{z}_3)  \\
\ & \quad + (\mbox{multi-particle contributions}),
\end{split}
\end{equation}
which fully reproduces the single-particle holomorphically singular contributions to the mixed-helicity multi-graviton OPEs \eqref{boundary-graviton-holomorphic}.  Note that the anti-holomorphic singularities can also be extracted from a bulk analysis but require a \emph{distinct} anti-holomorphic collinear limit.  By contrast, our boundary derivation recovers all terms from the direct application of a single formula \eqref{generalizedwick-1}.

\subsection{Multi-particle graviton OPEs from MHV amplitudes}
\label{subsec:bulk-multi-particle}

To calculate the contribution from composite operators to multi-particle OPEs from a bulk holomorphic collinear limit, we restrict our attention to MHV graviton amplitudes, for which there exist relatively simple recursion relations. Our analysis follows the same general logic as the treatment of MHV gluon amplitudes in \cite{Guevara:2024ixn} but involves the additional technical challenge of manipulating MHV graviton recursion relations, as opposed to the closed-form formulas for MHV gluon amplitudes.\footnote{The recently-discovered new recursion relation for MHV graviton amplitudes in \cite{Guevara:2025tsm} may provide a more slick derivation of the composite contribution than the one presented here.} For completeness, here we derive both single- and multi-particle contributions to the multi-particle OPE for MHV graviton amplitudes. 

MHV graviton amplitudes by definition have exactly two external negative-helicity gravitons and the remaining external particles are positive-helicity gravitons. Since the holomorphic (but not anti-holomorphic) singular terms in the multi-particle Einstein gravity OPE preserve the number of negative-helicity gravitons, we expect the expression \eqref{boundary-graviton-holomorphic} with only holomorphic singularities (as opposed to the more general mixed-helicity expressions in subsection \ref{subsec:general-ope}) to hold as an operator statement within the MHV sector. 

For this analysis, we need the little group scaling relation \cite{Elvang:2013cua}: 
\begin{equation} \label{little-group}
    \begin{split}
        \mathcal{A}_n \left( \{\lambda_1, \tilde \lambda_1 \}, \cdots, \{t_i \lambda_i, t_i^{-1} \tilde \lambda_i  \}, \cdots\right)
         = t_i^{-2s_i} \mathcal{A}_n\left( \{\lambda_1, \tilde \lambda_1 \}, \cdots, \{\lambda_i, \tilde \lambda_i  \}, \cdots\right).
    \end{split}
\end{equation}
Then, the central tool for this analysis is the recursion relation between  $n$- and $(n-1)$-point MHV graviton amplitudes \cite{Hodges:2011wm, Guevara:2019ypd} 
 \begin{equation} \label{MHV-recursion}
     \begin{split}
         \mathcal{A}_{n }^{\mbox{\scriptsize{MHV}}} & = \mathcal{A}_{n }^{\mbox{\scriptsize{MHV}}}(\{\lambda_1, \tilde{\lambda}_1\}, \ldots, \{\lambda_n, \tilde{\lambda}_n\}) \\
         &= \frac{\kappa}{2} \sum_{i=2}^{n-2} \frac{[1i] \langle ni \rangle \langle (n-1)i \rangle}{\langle 1i\rangle\langle n1 \rangle \langle (n-1)1 \rangle} \mathcal{A}_{n-1}^{\rm MHV} \left(\{\lambda_2, \tilde {\lambda}_{2} \},  \cdots, \left \{ \lambda_i, \tilde \lambda_i +\frac{\langle n1 \rangle}{\langle ni \rangle} \tilde{\lambda}_1 \right\}, \cdots, \left\{ \lambda_n, \tilde \lambda_n +\frac{\langle i1 \rangle}{\langle in \rangle} \tilde{\lambda}_1 \right\} \right).
     \end{split}
 \end{equation}
 Note that from the little-group scaling of the prefactor, it is straightforward to see that ``particle $1$'' must be a positive-helicity graviton.  Nevertheless, one can formulate recursion relations for a graviton of any helicity by working instead with the ``stripped'' amplitude $A_n^{\rm MHV}$, which for a scattering process with negative-helicity gravitons $i$ and $j$ is related to the ``unstripped''\footnote{Here stripped and unstripped refer to the factor of $\langle ij \rangle^8$, not the delta function in momentum.} amplitude by
 \begin{equation} \label{stripped-def}
     \begin{split}
          \mathcal{A}_{n }^{\mbox{\scriptsize{MHV}}}
           = \langle ij \rangle^8 A_n^{\rm MHV}, \quad \quad \quad s_i =s_j = -2.
     \end{split}
 \end{equation}
 Then the stripped amplitudes obey an analogous relation without making any assumption about the helicity of the removed particle:
  \begin{equation} \label{MHV-recursion-stripped}
     \begin{split}
         A_{n }^{\mbox{\scriptsize{MHV}}}  
         &= \frac{\kappa}{2} \sum_{i=2}^{n-2} \frac{[1i] \langle ni \rangle \langle (n-1)i \rangle}{\langle 1i\rangle\langle n1 \rangle \langle (n-1)1 \rangle} A_{n-1}^{\rm MHV} \left(\{\lambda_2, \tilde {\lambda}_{2} \},  \cdots,  \left\{ \lambda_i, \tilde \lambda_i +\frac{\langle n1 \rangle}{\langle ni \rangle} \tilde{\lambda}_1\right\}, \cdots, \left\{ \lambda_n, \tilde \lambda_n +\frac{\langle i1 \rangle}{\langle in \rangle} \tilde{\lambda}_1 \right\} \right).
     \end{split}
 \end{equation}

We begin by determining an explicit expression for a momentum space amplitude with a ``normal-ordered'' or composite pair of particles.  We introduce the following notation for the $\Theta\LP z_{ij}^0 \bar{z}_{ij}^0 \RP$ term in the expansion of the amplitude $\mathcal{A}_n$:
\begin{equation}
    \begin{split}
        \left[\mathcal{A}_n \right]_{:ij:} = \left[\mathcal{A}_n \right]_{:ij:} (\omega_i,\omega_j, z_j, \bar{z}_j),
    \end{split}
\end{equation}
where we have suppressed the dependence on other momentum $p_k$ for $k \neq i, j$.  In particular, the ``normal-ordered'' momentum space amplitude $\left[\mathcal{A}_n \right]_{:ij:}$ is defined, so that under a Mellin transform, the corresponding celestial amplitude contains the insertion of a normal-ordered celestial operator
\begin{equation} \label{mellin-transform-amp}
    \begin{split}
        \langle \cdots :G^{s_i}_{\Delta_i} G^{s_j}_{\Delta_j}: (z_j, \bar{z}_j) \cdots\rangle
         = \left(\prod_{k = 1}^n \int_0^\infty \frac{d \omega_k}{\omega_k} \omega_k^{\Delta_k}\right)\left[\mathcal{A}_n \right]_{:ij:}(\omega_i, \omega_j, z_j, \bar{z}_j). 
    \end{split}
\end{equation}

To find an explicit expression for normal-ordering $:12:$, we use the recursion relation \eqref{MHV-recursion} to expose the dependence on $\lambda_1, \tilde{\lambda}_1$ and then extract the $ \Theta \LP z_{12}^0\bar{z}_{12}^0 \RP$ term.  From the expression for the stripped amplitude \eqref{stripped-def} and the recursion \eqref{MHV-recursion-stripped}, it is straightforward to see that for $s_1=s_2=-2$, the amplitude vanishes in the limit $(z_{1}, \bar{z}_{1}) \to  (z_{2}, \bar{z}_{2})$ so
\begin{equation} \label{normal-order-negative}
    \begin{split}
         \left[\mathcal{A}_n^{\rm MHV} \right]_{:12:} = 0, \quad \quad s_1 = s_2= -2. 
    \end{split}
\end{equation}

To find an expression when $s_1 = +2$, we use \eqref{MHV-recursion}.  Noting that the $i = 2$ term scales like $\mathcal{O}(\bar{z}_{12})$ in the limit $1 \to 2$ and so will not contribute to the normal-ordered expression, we find 
\begin{equation} \label{normal-order-simple}
    \begin{split}
        &\left[ \mathcal{A}_{n}^{\rm MHV}  \right]_{:12:}^{s_1=+2}(\omega_1, \omega_2, z_2, \bar{z}_2)\\
        & \quad  = \frac{\kappa}{2}\frac{\omega_2}{\omega_1} \sum_{i=3}^{n-2} \frac{[2i] \langle ni \rangle \langle (n-1)i \rangle}{\langle 2i\rangle\langle n2 \rangle \langle (n-1)2\rangle} \mathcal{A}_{n-1}^{\rm MHV} \left(\{\lambda_2, \tilde {\lambda}_{2} \},  {\cdots}, \left  \{ \lambda_i, \tilde \lambda_i +\frac{\omega_1}{\omega_2}\frac{\langle n2 \rangle}{\langle ni \rangle} \tilde{\lambda}_2 \right\}, {\cdots},  \left \{ \lambda_n, \tilde \lambda_n +\frac{\omega_1}{\omega_2}\frac{\langle i2 \rangle}{\langle in \rangle} \tilde{\lambda}_2 \right\} \right).
    \end{split}
\end{equation}
Here the factors of $\omega_1/\omega_2$ arise from 
\begin{equation} \label{spinor-manip}
    \begin{split}
        \left.\left(\lambda_1, \tilde{\lambda}_1\right) \right|_{\substack{z_1=z_2\\ \bar{z}_1 = \bar{z}_2} } =\frac{\sqrt{\omega_1}}{\sqrt{\omega_2}}\left(\lambda_2, \tilde{\lambda}_2\right).
    \end{split}
\end{equation}
We also note that the $\Theta(z_{12}^0 \bar{z}_{12}^0)$ piece of $\mathcal{A}_{n}^{\rm MHV}$ after taking $m$ derivatives with respect to $\bar{z}_1$, which we denote with the subscript $:(\bar{\partial}^m1)2:$, is given by
\begin{equation} \label{normal-order-derivative-on-1}
    \begin{split}
        &\left[ \mathcal{A}_{n}^{\rm MHV}  \right]_{:(\bar{\partial}^m1)2:}^{s_1=+2}(\omega_1, \omega_2, z_2, \bar{z}_2)\\
        & ~~  = \frac{\kappa}{2}\frac{\omega_2}{\omega_1} \partial_{\bar{z}'_2}^m \left. \sum_{i=3}^{n-2} \frac{[2'i] \langle ni \rangle \langle (n{-}1)i \rangle}{\langle 2i\rangle\langle n2 \rangle \langle (n{-}1)2\rangle} \mathcal{A}_{n-1}^{\rm MHV} \left(\{\lambda_2, \tilde {\lambda}_{2} \},  {\cdots}, \left  \{ \lambda_i, \tilde \lambda_i {+}\frac{\omega_1}{\omega_2}\frac{\langle n2 \rangle}{\langle ni \rangle} \tilde{\lambda}'_2\right\}, {\cdots}, \left\{ \lambda_n, \tilde \lambda_n {+}\frac{\omega_1}{\omega_2}\frac{\langle i2 \rangle}{\langle in \rangle} \tilde{\lambda}'_2 \right\} \right) \right|_{2'=2}.
    \end{split}
\end{equation}
Similarly, the $\Theta(z_{12}^0 \bar{z}_{12}^0)$ piece of $\mathcal{A}_{n}^{\rm MHV}$ after taking $m$ derivatives with respect to $\bar{z}_2$, which we denote with the subscript $: 1(\bar{\partial}^m2):$, is given by 
\begin{equation} \label{normal-order-derivative-on-2}
    \begin{split}
        &\left[ \mathcal{A}_{n}^{\rm MHV}  \right]_{:1(\bar{\partial}^m2):}^{s_1=+2}(\omega_1, \omega_2, z_2, \bar{z}_2)\\
        & ~~  = \frac{\kappa}{2}\frac{\omega_2}{\omega_1} \partial_{\bar{z}_2}^m \left. \sum_{i=3}^{n-2} \frac{[2'i] \langle ni \rangle \langle (n{-}1)i \rangle}{\langle 2i\rangle\langle n2 \rangle \langle (n{-}1)2\rangle} \mathcal{A}_{n-1}^{\rm MHV} \left(\{\lambda_2, \tilde {\lambda}_{2} \},  {\cdots}, \left \{ \lambda_i, \tilde \lambda_i {+}\frac{\omega_1}{\omega_2}\frac{\langle n2 \rangle}{\langle ni \rangle} \tilde{\lambda}'_2 \right\}, {\cdots}, \left  \{ \lambda_n, \tilde \lambda_n {+}\frac{\omega_1}{\omega_2}\frac{\langle i2 \rangle}{\langle in \rangle} \tilde{\lambda}'_2 \right \} \right) \right|_{2'=2}.
    \end{split}
\end{equation}

Now we derive the $G^{+}_1:G_2^{+}G_3^{\pm}:$ OPE.  We begin with the $n$-point amplitude, this time normal-ordered with respect to $:23:$
\begin{equation}  \label{MHV-derivation-intermediate-1}
    \begin{split}
        &\left[ \mathcal{A}_{n}^{\rm MHV}  \right]^{s_2=+2}_{:23:}(\omega_2, \omega_3, z_3, \bar{z}_3)\\
        & \quad 
         = \frac{\kappa}{2}\frac{\omega_3}{\omega_2} \sum_{\substack{i\neq 2,3}}^{n-2} \frac{[3i] \langle ni \rangle \langle (n-1)i \rangle}{\langle 3i\rangle\langle n3 \rangle \langle (n-1)3\rangle} \mathcal{A}_{n-1}^{\rm MHV} \left(\{\lambda_1, \tilde {\lambda}_{1} \},\{\lambda_3, \tilde {\lambda}_{3} \},  {\cdots},  \left \{ \lambda_i, \tilde \lambda_i +\frac{\omega_2}{\omega_3}\frac{\langle n3 \rangle}{\langle ni \rangle} \tilde{\lambda}_3\right\}, \right. \\& \quad \quad \quad \quad \quad \quad 
        \quad \quad \quad \quad \quad \quad \quad \quad \quad \quad \quad \quad \quad \quad \quad \quad \quad \quad \quad \quad \quad \left. {\cdots}, \left \{ \lambda_n, \tilde \lambda_n +\frac{\omega_2}{\omega_3}\frac{\langle i3 \rangle}{\langle in \rangle} \tilde{\lambda}_3\right\} \right).
    \end{split}
\end{equation}
To take the limit $1\to 3$, we recurse one more time to expose the $1$ dependence of the amplitude.  We assume $s_1=+2$ so we can use \eqref{MHV-recursion} directly. For this purpose, it is helpful to separate \eqref{MHV-derivation-intermediate-1} into the $i=1$ and $i\neq 1$ terms.  We have
\begin{equation}
    \begin{split}
        &\left[ \mathcal{A}_{n}^{\rm MHV}  \right]^{s_1=s_2=+2}_{:23:}(\omega_2, \omega_3, z_3, \bar{z}_3) \Big|_{i=1}\\ 
        & \quad =  \frac{\kappa}{2}\frac{\omega_3}{\omega_2}\frac{[13]}{\langle 13\rangle}  
        \frac{\kappa}{2}\sum_{j = 3}^{n-2} \frac{[1j] + \frac{\omega_2}{\omega_3}\frac{\langle n3 \rangle}{\langle n1 \rangle}[3j]}{\langle 1j\rangle} \frac{\langle nj \rangle \langle (n-1)j\rangle}{\langle n3 \rangle \langle (n-1)3\rangle}\\& 
        \quad  \times \mathcal{A}_{n-2}^{\rm MHV} \left(\{\lambda_3, \tilde {\lambda}_{3} \},  \cdots, \left  \{\lambda_j, \tilde{\lambda}_j + \frac{\langle n 1\rangle}{\langle nj \rangle }\tilde {\lambda}_{1} +\frac{\omega_2}{\omega_3}\frac{\langle n3 \rangle} {\langle nj \rangle } \tilde{\lambda}_3 \right \},\cdots,    \left\{ \lambda_n, \tilde \lambda_n +   \frac{\langle j1 \rangle}{\langle jn \rangle}\tilde {\lambda}_{1} +\frac{\omega_2}{\omega_3}\frac{\langle j3 \rangle}{\langle jn \rangle} \tilde{\lambda}_3 \right \} \right)
    \end{split}
\end{equation}
and 
\begin{equation}
    \begin{split}
         &\left[ \mathcal{A}_{n}^{\rm MHV}  \right]^{s_1=s_2=+2}_{:23:}(\omega_2, \omega_3, z_3, \bar{z}_3) \Big|_{i\neq1}\\ 
         &=\frac{\kappa}{2}\frac{\omega_3}{\omega_2} \sum_{i=4}^{n-2} \frac{[3i] \langle ni \rangle \langle (n-1)i \rangle}{\langle 3i\rangle\langle n3 \rangle \langle (n-1)3\rangle}\frac{\kappa}{2}
         \sum_{j=3}^{n-2}\frac{[1j]}{\langle 1j\rangle} \frac{\langle nj \rangle \langle (n-1)j \rangle }{\langle n1 \rangle \langle (n-1)1 \rangle }\\& ~~    {\times}  
         \mathcal{A}_{n-2}^{\rm MHV} \Bigg( \{\lambda_3, \tilde {\lambda}_{3} \}, {\cdots},  \left \{ \lambda_i, \tilde \lambda_i {+}\frac{\omega_2}{\omega_3}\frac{\langle n3 \rangle}{\langle ni \rangle} \tilde{\lambda}_3 \right \}, {\cdots},  \left \{ \lambda_j, \tilde \lambda_j {+}  \frac{\langle n1 \rangle}{\langle nj \rangle} \tilde{\lambda}_1 \right \}, {\cdots}, \left \{ \lambda_n, \tilde \lambda_n {+}\frac{\omega_2}{\omega_3}\frac{\langle i3 \rangle}{\langle in \rangle} \tilde{\lambda}_3{+} \frac{\langle j1\rangle}{\langle jn\rangle }\tilde{\lambda}_1\right\} \Bigg),
    \end{split}
\end{equation}
respectively. For the $i =1$ contribution, it is helpful to further separate out the $j=3$ term.  Then, we find
\begin{equation}
    \begin{split}
        &\left[ \mathcal{A}_{n}^{\rm MHV}  \right]^{s_1=s_2=+2}_{:23:}(\omega_2, \omega_3, z_3, \bar{z}_3) \Big|_{i=1}\\
        &=  \frac{\kappa^2}{4}\frac{\omega_3}{\omega_2}  \frac{[13]^2 }{\langle 13\rangle^2}  \mathcal{A}_{n-2}^{\rm MHV} \left( \left \{\lambda_3, \frac{\omega_2+\omega_3}{\omega_3}\tilde{\lambda}_3 + \frac{\langle n 1\rangle}{\langle n3 \rangle }\tilde {\lambda}_{1}  \right\},\cdots,   \left \{ \lambda_n, \tilde \lambda_n +   \frac{\langle 31 \rangle}{\langle 3n \rangle}\tilde {\lambda}_{1}  \right \} \right)\\
        & \quad  +\frac{\kappa}{2}\frac{\omega_3}{\omega_2}\frac{[13]}{\langle 13\rangle}  
        \frac{\kappa}{2}\sum_{j = 4}^{n-2} \frac{[1j] + \frac{\omega_2}{\omega_3}\frac{\langle n3 \rangle}{\langle n1 \rangle}[3j]}{\langle 1j\rangle} \frac{\langle nj \rangle \langle (n-1)j\rangle}{\langle n3 \rangle \langle (n-1)3\rangle}\\& 
        \quad \quad   \times \mathcal{A}_{n-2}^{\rm MHV} \left(\{\lambda_3, \tilde {\lambda}_{3} \},  \cdots, \left  \{\lambda_j, \tilde{\lambda}_j + \frac{\langle n 1\rangle}{\langle nj \rangle }\tilde {\lambda}_{1} +\frac{\omega_2}{\omega_3}\frac{\langle n3 \rangle} {\langle nj \rangle } \tilde{\lambda}_3 \right  \},\cdots,     \left \{ \lambda_n, \tilde \lambda_n +   \frac{\langle j1 \rangle}{\langle jn \rangle}\tilde {\lambda}_{1} +\frac{\omega_2}{\omega_3}\frac{\langle j3 \rangle}{\langle jn \rangle} \tilde{\lambda}_3 \right \} \right).
    \end{split}
\end{equation}
Next we expand in the limit $z_1 \to z_3$. We work at fixed $\bar{z}_1$ and in the next step will expand to all orders in the difference $\bar{z}_{13}$.  This subsequent expansion in $\bar{z}_{13}$ corresponds to the derivative expansion on the right-hand side of the multi-particle OPE. To express the result in a compact and intelligible form, we use the explicit relations in appendix \ref{sec:conventions} to simplify spinor expressions, such as for example
\begin{equation}
    \begin{split}
        \frac{\omega_2+\omega_3}{\omega_3}\tilde{\lambda}_3 + \frac{\langle n 1\rangle}{\langle n3 \rangle }\tilde {\lambda}_{1} 
        &= \frac{\omega_1+\omega_2+\omega_3}{\omega_3} \tilde \lambda \left(\omega_3, \bar{z}_3 + \frac{\omega_1 \bar{z}_{13}}{\omega_1+\omega_2+\omega_3} \right)  + \mathcal{O}(z_{13}). 
    \end{split}
\end{equation}
Then, defining
\begin{equation}
    \begin{split}
        \tilde \lambda'_3 &\equiv \tilde \lambda \left(\omega_3, \bar{z}_3 + \frac{\omega_1 \bar{z}_{13}}{\omega_1+\omega_2+\omega_3} \right), \quad \quad \quad 
        \tilde \lambda''_3   \equiv  \tilde \lambda \left(\omega_3, \bar{z}_3 + \frac{\omega_1 \bar{z}_{13}}{\omega_1+\omega_2} \right),
    \end{split}
\end{equation}
we find  
\begin{equation}
    \begin{split}
        &\left[ \mathcal{A}_{n}^{\rm MHV}  \right]^{s_1=s_2=+2}_{:23:}(\omega_2, \omega_3, z_3, \bar{z}_3) \Big|_{i=1}\\
        &\sim   \frac{\kappa^2}{4}\frac{\omega_3}{\omega_2}  \frac{\bar{z}_{13}^2 }{z_{13}^2}  \mathcal{A}_{n-2}^{\rm MHV} \left( \left \{\lambda_3,\frac{\omega_1+\omega_2+\omega_3}{\omega_3} \tilde \lambda'_3 \right\},\cdots,     \{ \lambda_n, \tilde \lambda_n  \} \right)\\
        & \quad  -\frac{\kappa}{2}\frac{\omega_3 }{\omega_2}\frac{\omega_1+\omega_2}{\omega_1}\frac{\bar{z}_{13}}{z_{13}} \frac{\kappa}{2}\sum_{j = 4}^{n-2} \frac{  [3''j] }{ \langle 3j\rangle} \frac{\langle nj \rangle \langle (n-1)j\rangle}{\langle n3 \rangle \langle (n-1)3\rangle}\\& 
        \quad \quad   \times \mathcal{A}_{n-2}^{\rm MHV} \left(\{\lambda_3, \tilde {\lambda}_{3} \},  \cdots,  \left \{\lambda_j, \tilde{\lambda}_j + \frac{\langle n 3\rangle}{\langle nj \rangle} \frac{\omega_1+\omega_2}{\omega_3} \tilde \lambda''_{3}  \right\},\cdots,  \left\{ \lambda_n, \tilde \lambda_n +   \frac{\langle j3 \rangle}{\langle jn \rangle} \frac{\omega_1+\omega_2}{\omega_3} \tilde \lambda''_{3}  \right\} \right).
    \end{split}
\end{equation}

To re-express the amplitude in terms of standard momentum-space celestial variables $\omega_i,z_i, \bar{z}_i$, we need to perform a little group scaling.  Namely, we use \eqref{little-group} to write
\begin{equation}
    \begin{split}
        &\mathcal{A}_{n-2}^{\rm MHV} \left(  \left \{\lambda_3, \frac{\omega_1+\omega_2+\omega_3}{\omega_3} \tilde \lambda'_3   \right\},\cdots,    \{ \lambda_n, \tilde \lambda_n  \} \right)\\
         &~~= \frac{(\omega_1{+}\omega_2{+}\omega_3)^{s_3}}{\omega_3^{s_3}}
          \underbrace{\mathcal{A}_{n-2}^{\rm MHV} \left( \left \{\sqrt{\frac{\omega_1{+}\omega_2{+}\omega_3}{\omega_3}}\lambda_3, \sqrt{\frac{\omega_1{+}\omega_2{+}\omega_3}{\omega_3}}  \tilde \lambda'_3 \right\},\cdots,    \{ \lambda_n, \tilde \lambda_n  \} \right)}_{
         = \mathcal{A}_{n-2}^{\rm MHV} \LP \left \{\omega_1+\omega_2+\omega_3, z_{3},\bar{z}_3 +  \frac{\omega_1\bar{z}_{13}}{\omega_1 + \omega_2 + \omega_3} \right\}, \cdots,  \{ \lambda_n, \tilde \lambda_n  \} \RP}.
    \end{split}
\end{equation}
This $j = 3$ term then naturally admits a Taylor expansion in $\bar{z}_{13}$.

Next, we observe that if $\tilde \lambda''_3 = \tilde \lambda_3$, then the $j\neq3$ terms organize into the form of a normal-ordered amplitude \eqref{normal-order-simple}.  Moreover, the expansion of $\tilde \lambda''_3$ about $\tilde \lambda_3$, which is equivalently an expansion in $\bar{z}_{13}$ produces normal-ordered derivatives of amplitudes of the form \eqref{normal-order-derivative-on-1}.  Putting this all together, we find the $i=1$ term gives the simplified contribution:
 \begin{equation} \label{final-i=1term}
     \begin{split}
         & \left[ \mathcal{A}_{n}^{\rm MHV}  \right]^{s_1=s_2=+2}_{:23:}(\omega_2, \omega_3, z_3, \bar{z}_3) \Big|_{i=1} \\
        & \sim  \frac{\kappa^2}{4} \frac{\bar{z}_{13}^2}{z_{13}^2} \frac{(\omega_1+\omega_2+\omega_3)^{s_3}}{\omega_2 \omega_3^{s_3-1}} \sum_{m=0}^\infty \frac{\bar{z}_{13}^{m}}{m!}\frac{ \omega_1^m}{(\omega_1+\omega_2+\omega_3)^{m}} \D^m_{\bar{z}_3} \mathcal{A}_{n-2}^{\rm MHV} (\{\omega_1+\omega_2+\omega_3, z_{3},\bar{z}_3\}, \cdots )\\
        & \quad -\frac{\kappa}{2} \frac{\bar{z}_{13}}{z_{13}} \frac{(\omega_1+\omega_2)^2}{\omega_1\omega_2} \sum_{m=0}^\infty \frac{\bar{z}^{m}_{13}}{m!}   \frac{ \omega_1^{m}}{(\omega_1+\omega_2)^{m}} 
        \left[ \mathcal{A}_{n-1}^{\rm MHV}\right]^{s_{(1+2)}=+2}_{: \LP \bar{\D}^m (1+2) \RP 3:}(\omega_1+\omega_2, \omega_3, z_{3}, \bar{z}_3).
     \end{split}
 \end{equation}

Finally, we consider the $i\neq 1$ contribution, again taking $z_1 \to z_3$ at fixed $\bar{z}_1$.  The only singular terms in $z_{13}$ arise from the $j =3$ term, so we find
 \begin{equation} \label{ineqj-int1}
    \begin{split}
         &\left[ \mathcal{A}_{n}^{\rm MHV}  \right]^{s_1=s_2=+2}_{:23:}(\omega_2, \omega_3, z_3, \bar{z}_3) \Big|_{i\neq1}\\ 
        & \quad \sim -\frac{\kappa}{2}\frac{\omega_3^2}{\omega_1\omega_2} \frac{\bar{z}_{13}}{z_{13}}\frac{\kappa}{2}\sum_{i=4}^{n-2} \frac{[3i] \langle ni \rangle \langle (n-1)i \rangle}{\langle 3i\rangle\langle n3 \rangle \langle (n-1)3\rangle} \\& \quad \quad    \times  
         \mathcal{A}_{n-2}^{\rm MHV} \left( \left\{ \lambda_3,   \frac{\omega_1+\omega_3}{\omega_3} \tilde{\lambda}'''_3\right\},\cdots, \left \{ \lambda_i, \tilde \lambda_i +\frac{\omega_2}{\omega_3}\frac{\langle n3 \rangle}{\langle ni \rangle} \tilde{\lambda}_3\right\}, \cdots, \left  \{ \lambda_n, \tilde \lambda_n +\frac{\omega_2}{\omega_3}\frac{\langle i3 \rangle}{\langle in \rangle} \tilde{\lambda}_3\right\} \right), 
    \end{split}
\end{equation}
where here
\begin{equation}
    \begin{split}
        \tilde \lambda'''_3 = \tilde \lambda \left(\omega_3, \bar{z}_3 + \frac{\omega_1}{\omega_1 + \omega_3} \bar{z}_{13}\right).
    \end{split}
\end{equation}
In order to identify the sum over $(n-2)$-point amplitudes with a normal-ordered amplitude as in \eqref{normal-order-simple}, we must perform a little group scaling on particle 3 and write the answer in terms of 
\begin{equation}
    \begin{split}
        \hat \lambda_3  \equiv \sqrt{\frac{\omega_1+\omega_3}{\omega_3}} \lambda_3, \quad \quad 
         \hat{\tilde{\lambda}}_3  \equiv \sqrt{\frac{\omega_1+\omega_3}{\omega_3}} \tilde{\lambda} _3, \quad \quad \quad \hat{\tilde{\lambda}}'''_3  \equiv \sqrt{\frac{\omega_1+\omega_3}{\omega_3}} \tilde{\lambda}''' _3.
    \end{split}
\end{equation}
Doing this, we find
\begin{equation}
    \begin{split}
      & \mathcal{A}_{n-2}^{\rm MHV}\left( \left\{ \lambda_3,   \frac{\omega_1+\omega_3}{\omega_3}  \tilde{\lambda}'''_3 \right\},\cdots,  \left\{ \lambda_i, \tilde \lambda_i +\frac{\omega_2}{\omega_3}\frac{\langle n3 \rangle}{\langle ni \rangle} \tilde{\lambda}_3 \right\}, \cdots,  \left \{ \lambda_n, \tilde \lambda_n +\frac{\omega_2}{\omega_3}\frac{\langle i3 \rangle}{\langle in \rangle} \tilde{\lambda}_3\right \} \right)\\
         &\quad  =  \frac{(\omega_1+\omega_3)^{s_3}}{ \omega_3^{s_3}}
            \mathcal{A}_{n-2}^{\rm MHV} \left( \{ \hat \lambda_3,    \hat{\tilde{\lambda}}'''_3 \},\cdots, \left \{ \lambda_i, \tilde \lambda_i +\frac{\omega_2}{\omega_1+\omega_3}\frac{\langle n\hat{3} \rangle}{\langle ni \rangle} \hat{\tilde{\lambda}}_3\right\},   \cdots,  \left \{ \lambda_n, \tilde \lambda_n +\frac{\omega_2}{\omega_1+\omega_3}\frac{\langle i \hat{3} \rangle}{\langle in \rangle} \hat{\tilde{\lambda}}_3\right \} \right),
    \end{split}
\end{equation}
and substituting back into our expression \eqref{ineqj-int1}, we find 
 \begin{equation}
    \begin{split}
         &\left[ \mathcal{A}_{n}^{\rm MHV}  \right]^{s_1=s_2=+2}_{:23:}(\omega_2, \omega_3, z_3, \bar{z}_3) \Big|_{i\neq1}\\ 
         & \quad \sim -\frac{\kappa}{2}\frac{\omega_3^2}{\omega_1\omega_2} 
          \frac{(\omega_1+\omega_3)^{s_3+1}}{ \omega_3^{s_3+1}}\frac{\bar{z}_{13}}{z_{13}}\frac{\kappa}{2}\sum_{i=4}^{n-2} \frac{[\hat 3i] \langle ni \rangle \langle (n-1)i \rangle}{\langle \hat 3i\rangle\langle n \hat 3 \rangle \langle (n-1) \hat 3\rangle} \\& \quad \quad    \times  
            \mathcal{A}_{n-2}^{\rm MHV} \left( \{ \hat \lambda_3,    \hat{\tilde{\lambda}}'''_3 \},\cdots,  \left\{ \lambda_i, \tilde \lambda_i +\frac{\omega_2}{\omega_1+\omega_3}\frac{\langle n\hat{3} \rangle}{\langle ni \rangle} \hat{\tilde{\lambda}}_3\right\},   \cdots,  \left \{ \lambda_n, \tilde \lambda_n +\frac{\omega_2}{\omega_1+\omega_3}\frac{\langle i \hat{3} \rangle}{\langle in \rangle} \hat{\tilde{\lambda}}_3 \right\} \right).
    \end{split}
\end{equation}
Notice that when $\hat{\tilde{\lambda}}'''_3=\hat{\tilde{\lambda}}_3$, this reproduces the form of the normal-ordered amplitude \eqref{normal-order-simple} and when $\hat{\tilde{\lambda}}'''_3$ is expanded around $\hat{\tilde{\lambda}}_3$, each term reproduces the normal-ordered derivative expression in \eqref{normal-order-derivative-on-2}. This contribution thus assembles into the expression for a normal-ordered insertion \eqref{normal-order-simple} and its derivatives \eqref{normal-order-derivative-on-2} and therefore can be written as  
 \begin{equation}
    \begin{split}
        & \left[ \mathcal{A}_{n}^{\rm MHV}  \right]^{s_1=s_2=+2}_{:23:}(\omega_2, \omega_3, z_3, \bar{z}_3) \Big|_{i\neq1}  
            \\
            &\quad \quad \quad\sim -\frac{\kappa}{2} 
          \frac{(\omega_1+\omega_3)^{s_3}}{\omega_1 \omega_3^{s_3-1}} \frac{\bar{z}_{13}}{z_{13}}\sum_{m=0}^\infty \frac{\bar{z}^{m}_{13}}{m!}   \frac{ \omega_1^{m } }{ (\omega_1+\omega_3)^{m}}   \left[ \mathcal{A}_{n-1}^{\rm MHV}\right]^{s_2=+2}_{:2 (\bar{\partial}^m(1+3)):}(\omega_2, \omega_1+\omega_3, z_{3},\bar{z}_3).
    \end{split}
\end{equation}
Combining this result with the terms from $i=1$ \eqref{final-i=1term}, we find the following momentum space expression for the $z_1 \to z_3$ limit of the amplitude with $:23:$ normal-ordering: 
 \begin{equation}  
     \begin{split}
         & \left[ \mathcal{A}_{n}^{\rm MHV}  \right]^{s_1=s_2=+2}_{:23:}(\omega_2, \omega_3, z_3, \bar{z}_3) \\
         &  \quad \sim  \frac{\kappa^2}{4} \frac{\bar{z}_{13}^2}{z_{13}^2} \frac{(\omega_1+\omega_2+\omega_3)^{s_3}}{\omega_2 \omega_3^{s_3-1}} \sum_{m=0}^\infty \frac{\bar{z}_{13}^{m}}{m!}\frac{ \omega_1^m}{(\omega_1+\omega_2+\omega_3)^{m}} \D^m_{\bar{z}_3} \mathcal{A}_{n-2}^{\rm MHV} (\{\omega_1+\omega_2+\omega_3, z_{3},\bar{z}_3\}, \cdots )\\
        & \quad \quad -\frac{\kappa}{2} \frac{\bar{z}_{13}}{z_{13}} \frac{(\omega_1+\omega_2)^2}{\omega_1\omega_2} \sum_{m=0}^\infty \frac{\bar{z}^{m}_{13}}{m!}   \frac{ \omega_1^{m}}{(\omega_1+\omega_2)^{m}} 
        \left[ \mathcal{A}_{n-1}^{\rm MHV}\right]^{s_{(1+2)}=+2}_{: \LP \bar{\D}^m (1+2) \RP 3:}(\omega_1+\omega_2, \omega_3, z_{3}, \bar{z}_3)
          \\
        &\quad  \quad -\frac{\kappa}{2} 
          \frac{(\omega_1+\omega_3)^{s_3}}{\omega_1 \omega_3^{s_3-1}} \frac{\bar{z}_{13}}{z_{13}}\sum_{m=0}^\infty \frac{\bar{z}^{m}_{13}}{m!}   \frac{ \omega_1^{m } }{ (\omega_1+\omega_3)^{m}}   \left[ \mathcal{A}_{n-1}^{\rm MHV}\right]^{s_2=+2}_{:2 (\bar{\partial}^m(1+3)):}(\omega_2, \omega_1+\omega_3, z_{3},\bar{z}_3).
     \end{split}
 \end{equation} Then, using \eqref{mellin-transform-amp}, the corresponding celestial amplitudes are found to obey the following relation
 \begin{equation}
     \begin{split}
         & \langle G^+_{\Delta_1}(z_1, \bar{z}_1) : G^+_{\Delta_2}G^\pm_{\Delta_3}:(z_3, \bar{z}_3) \cdots\rangle \\
         & \quad \sim \frac{\kappa^2}{4} \frac{\bar{z}_{13}^2}{z_{13}^2}\sum_{m=0}^\infty \frac{\bar{z}_{13}^{m}}{m!}   B(2 \bar{h}_1 + 2 + m, 2 \bar{h}_2 + 1, 2 \bar{h}_3 + 1) \langle \bar{\partial}^m G^\pm_{\Delta_1+\Delta_2+\Delta_3} (z_3, \bar{z}_3) \cdots \rangle\\
         &\quad \quad - \frac{\kappa}{2} \frac{\bar{z}_{13}}{z_{13}}\sum_{m=0}^\infty \frac{\bar{z}^{m}_{13}}{m!}  B(2 \bar{h}_1+1 + m, 2 \bar{h}_2 + 1) \langle  :\LP \bar{\D}^m G^{+} _{\Delta_1 + \Delta_2} \RP G_{\Delta_3}^{\pm }: (z_3, \bar{z}_3)\cdots \rangle \\
         & \quad \quad - \frac{\kappa}{2} \frac{\bar{z}_{13}}{z_{13}}\sum_{m=0}^\infty \frac{\bar{z}^{m}_{13}}{m!}B(2 \bar{h}_1 + 1 + m, 2 \bar{h}_3 + 1) \langle  :G_{\Delta_2}^{+} \bar{\D}^m G^{\pm}_{\Delta_1 + \Delta_3} :(z_3, \bar{z}_3)\cdots \rangle,
     \end{split}
 \end{equation}
where here as above we have kept only the singular terms in the limit $z_1 \to z_3$ while retaining all orders in the expansion of $\bar{z}_1$ about $\bar{z}_3$.  Comparing with \eqref{boundary-graviton-holomorphic}, we exactly recover the multi-graviton OPE that was previously computed by boundary methods. 

The multi-graviton OPEs in \eqref{boundary-graviton-holomorphic} for other choices of bulk helicities $s_1$, $s_2$, $s_3$ can also be recovered from similar or simpler calculations.  In particular, note that if at least two of the gravitons $1$, $2$, or $3$ are negative helicity, then the right-hand side of \eqref{boundary-graviton-holomorphic} will vanish when inserted in an MHV graviton amplitude due to \eqref{normal-order-negative}.  The left-hand side will likewise vanish, because the negative helicity prefactor $\langle ij \rangle^8$ in \eqref{stripped-def} involves $i,j$ equal to some pair of $1,2,3$. Then, the limit $2 \to 3$ and $1 \to 3$, will eventually set all $\lambda_i$, $i=1,2,3$ to be parallel and since the most singular divergence of the stripped amplitudes is $\propto 1/\langle ij \rangle^2$, the result will vanish in the multi-collinear limit.  Finally, the other combinations of $s_1$, $s_2$, $s_3$ with two positive and one negative helicity can be derived by slightly modified versions of the analysis presented above.

\section{Multi-particle OPEs from symmetry}
\label{sec:symmetry}

In this section, we present a third independent method for determining the OPE coefficients in multi-particle OPEs. Like in section \ref{sec:boundary-derivation}, this derivation proceeds entirely within the boundary theory, but rather than specifying the single-particle OPEs, here we consider the implications of the celestial image of bulk translational symmetry, in addition to those of boundary global conformal symmetry ($\simeq$ the bulk proper orthochronous Lorentz symmetry).  Our strategy involves proposing a general ansatz for the multi-particle OPE and then constraining its form from symmetry.  The only input in this analysis is the transformation of both single- and multi-particle celestial operators under 4D bulk Poincar\'e symmetry. To deduce these transformations, specifically the transformations of multi-particle operators, we apply some of the methods from section \ref{sec:boundary-derivation}.  However, the logic underlying this symmetry-based method is conceptually distinct from that in section \ref{sec:boundary-derivation}.  Explicitly, in principle, under this symmetry-based method, one simply begins with a collection of operators with specified transformation properties under the enhanced symmetry group and then proceeds to constrain their OPEs.  In particular, from this perspective the identification of ``single-particle'' versus composite operators need not necessarily be given as input. Rather in this approach, these two classes of operators are distinguished simply by their respective transformations under symmetry.

More concretely, our analysis is a direct extension of that in \cite{Himwich:2021dau}. In this section, we need slightly more explicit notation for the single-particle celestial conformal primaries $\mathcal{O}_i$, which henceforth we label by the left and right conformal weights $\mathcal{O}_{h_i, \bar{h}_i}$. Then, given a generic  form\footnote{Apart from specifying the holomorphic structure to take the form of a simple pole. Note however, even this form of the ansatz can be derived from a symmetry perspective as was done for example in \cite{Banerjee:2020zlg,Banerjee:2020vnt,Banerjee:2025oyu}.} of an OPE between global conformal primaries that is consistent with boundary scaling, rotations and translations  
\begin{equation}\label{single-single-ansatz}
\begin{aligned}
\mc{O}_{h_1, \bar{h}_1} (z, \bar{z}) \mc{O}_{h_2, \bar{h}_2} (0, 0) & = \frac{1}{z} \sum_{p\geq 0}  \sum_{m=0}^\infty C_p^{(m)}(\bar{h}_1, \bar{h}_2) \bar{z}^{p + m} \bar{\D}^m \mc{O}_{h_1+h_2-1, \bar{h}_1+\bar{h}_2+p}(0,0)  + \mc{O} \LP z^0 \RP,
\end{aligned}
\end{equation}
it was shown in \cite{Himwich:2021dau} that covariance under the special conformal generator $\left[ \bar{L}_1, \cdot \right]$ (corresponding in the bulk to a certain Lorentz generator) implies the following recursion relation for OPE coefficients 
\begin{equation} \label{single-particle-m-recursion}
\LP 2\bar{h}_1 +p+ m  \RP C_p^{(m)} (\bar{h}_1, \bar{h}_2) = (m +1)(2\bar{h}_1 + 2\bar{h}_2 + 2p + m) C_p^{(m+1)}(\bar{h}_1, \bar{h}_2).
\end{equation} 
In the ansatz \eqref{single-single-ansatz}, $p$ is just a parameter that specifies the spin of the operator on the right-hand side $s_I = s_1+s_2-p-1$.  Hence the sum over $p$ is equivalent to the sum over possible conformal spin of a single-particle celestial primary that appeared in previous sections.   

Likewise, enforcing covariance under bulk translations further implies \cite{Himwich:2021dau} \begin{equation}
C^{(m)}_p\LP \bar{h}_1 + \frac{1}{2}, \bar{h}_2 \RP + C^{(m)}_p\LP \bar{h}_1, \bar{h}_2 + \frac{1}{2} \RP = C^{(m)}_p\LP \bar{h}_1, \bar{h}_2 \RP 
\end{equation} 
and \begin{equation}
C^{(m)}_p\LP \bar{h}_1 + \frac{1}{2}, \bar{h}_2 \RP = (m+1) C^{(m+1)}_p\LP \bar{h}_1, \bar{h}_2 \RP.
\end{equation}
Taking appropriate linear combinations of these constraints at $m = 0$ fixes $C^{(0)}_p(\hb_1, \hb_2)$, and subsequently applying the recursive constraint in $m$ \eqref{single-particle-m-recursion} fixes the general-$m$ coefficient to take the form 
\begin{equation}
C^{(m)}_p(\bar{h}_1, \bar{h}_2) = \frac{\gamma_{s_I}^{s_1, s_2}}{m!} B( 2\bar{h}_1 + p+ m, 2\bar{h}_2+p),
\end{equation}
where $s_I = s_1+s_2-p-1$. Note that this is precisely the single-particle OPE \eqref{general-single-particle-OPE} that we used as input for the analysis in subsection \ref{subsec:holomorphic-allspin}.

In this section, we extend the analysis of \cite{Himwich:2021dau} to the multi-particle OPE and in particular reproduce the form derived in subsection \ref{subsec:holomorphic-allspin}. To perform this analysis, we need the transformation of multi-particle celestial operators under bulk Poincar\'e symmetry.  We determine the ${\rm SL}(2, \mathbb{C})$ transformation properties in subsection \ref{subsec:sl2c} and the transformation under bulk translations in subsection \ref{subsec:translations}.  Analogues of these transformation rules in the Carrollian basis were previously found in \cite{Kulp:2024scx}. Then, in subsection \ref{subsec:symmetry-single-particle}, we present an ansatz for the single-particle contribution to the multi-particle OPE and determine the coefficients from symmetry.  In subsection \ref{subsec:symmetry-multi-particle} we treat the multi-particle contributions, again beginning with an ansatz and then fixing the coefficients from symmetry. 

\subsection{${\rm SL}(2, \mathbb{C})$ transformations of multi-particle operators}
\label{subsec:sl2c}

In this subsection, we determine the transformation properties of multi-particle celestial operators under ${\rm SL}(2, \mathbb{C})$. As we will see, these  transformation properties can be deduced from the symmetry transformations of the single-particle operators and the single-particle OPEs. 

In a 2D CFT, the  transformation of an operator under global conformal symmetry is determined by its operator product expansion with the stress tensor
\begin{equation} \label{def-sl2c-trans}
    \begin{split}
        \left[L_n, \mathcal{O}(z, \bar{z}) \right] = \oint_{z} \frac{dw}{2 \pi i } w^{n+1} T(w) \mathcal{O}(z, \bar{z}), \quad \quad \quad n = 0, \pm 1. 
    \end{split}
\end{equation}
Conformal primary operators admit an OPE of the form 
\begin{equation} \label{T-primary}
    \begin{split}
       T(w) \mathcal{O}_{h, \bar{h}}(z, \bar{z}) \sim \frac{h}{(w-z)^2} \mathcal{O}_{h, \bar{h}}(z,\bar{z}) + \frac{1}{w-z} \partial \mathcal{O}_{h, \bar{h}}(z,\bar{z})
    \end{split}
\end{equation}
and thus transform according to 
\begin{equation}
    \begin{split}
         \left[L_n, \mathcal{O}_{h, \bar{h}}(z, \bar{z}) \right] = z^n \left((n+1) h + z \partial \right)\mathcal{O}_{h, \bar{h}}(z,\bar{z}). 
    \end{split}
\end{equation}

To evaluate \eqref{def-sl2c-trans} for multi-particle operators, we need their OPEs with the stress tensor.  These can be calculated using the generalized Wick formula \eqref{generalizedwick-1} from section \ref{sec:boundary-derivation}
\begin{equation}
    \begin{split}
        T(z_0)&: \mc{O}_{h_1, \bar{h}_1} \mc{O}_{h_2, \bar{h}_2}:(z_2, \bar{z}_2)
           \\& \sim \oint_{z_2} \frac{d z_1}{2 \pi i  } \oint_{\bar{z}_{2}} \frac{d \bar{z}_1}{2 \pi i  } \frac{1}{z_{12}\bar{z}_{12}}\Bigg[ \LP \frac{h_1 \mc{O}_{h_1, \bar{h}_1} (z_1, \bar{z}_1)}{z_{01}^2} + \frac{\D \mc{O}_{h_1, \bar{h}_1} (z_1, \bar{z}_1)}{z_{01}} \RP \mc{O}_{h_2, \bar{h}_2}(z_2, \bar{z}_2) \\
& \quad \quad \quad \quad \quad \quad \quad  \quad \quad \quad \quad  + \mc{O}_{h_1, \bar{h}_1}(z_1, \bar{z}_1) \LP \frac{h_2 \mc{O}_{h_2, \bar{h}_2}(z_2, \bar{z}_2)}{z_{02}^2} + \frac{\D \mc{O}_{h_2, \bar{h}_2} (z_2, \bar{z}_2)}{z_{02}} \RP  \Bigg].
    \end{split}
\end{equation}
Note that the singular terms in the OPE will be sufficient to determine the transformation under global conformal symmetry. 

In general, this OPE is sensitive to singular terms in the $ \mc{O}_{h_1, \bar{h}_1} \mc{O}_{h_2, \bar{h}_2}$ OPE that give rise to single-particle contributions to the $T:\mathcal{O}_1\mathcal{O}_2:$ OPE.  However, from the form of the single-particle OPE \eqref{general-single-particle-OPE}, we observe that these corrections only contribute to the $T:\mathcal{O}_1\mathcal{O}_2:$ OPE when $12 \to I$ interactions have $s_I = s_1+s_2-1$. It is straightforward to see that the three-gluon coupling in pure Yang-Mills is an example of an interaction that satisfies $s_I = s_1+s_2-1$.  More generally, from \eqref{spin-constraint}, interactions with $s_I = s_1+s_2-1$ arise from bulk three-point vertices with $d_V = 4$.  Although of course, we are generally interested in theories that include this type of interaction, in this work we focus on the symmetry transformations and constraints that follow in the absence of these interactions. A proper treatment is beyond the scope of this work for reasons we will describe shortly. 

Assuming that no such $d_V=4$ interactions are present in our theory, or restricting to theories with only irrelevant bulk three-point vertices, we find 
\begin{equation}
    \begin{split}
        T(w) : \mc{O}_{h_1, \bar{h}_1} \mc{O}_{h_2, \bar{h}_2}:(z, \bar{z}) \sim \frac{h_1+h_2}{(w-z)^2} : \mc{O}_{h_1, \bar{h}_1} \mc{O}_{h_2, \bar{h}_2}:(z,\bar{z}) + \frac{1}{w-z} \partial : \mc{O}_{h_1, \bar{h}_1} \mc{O}_{h_2, \bar{h}_2}:(z,\bar{z}).
    \end{split}
\end{equation}
Note this is precisely the same form as the OPE for a conformal primary \eqref{T-primary} and accordingly, these composite operators transform under ${\rm SL}(2, \mathbb{C})$ like global conformal primaries
\begin{equation} \label{Ln}
\begin{aligned}
\left[L_n, :\mc{O}_{h_1, \bar{h}_1} \mc{O}_{h_2, \bar{h}_2}:(z , \bar{z} )\right] & = z^n \left((n+1) (h_1+h_2) + z \partial \right): \mc{O}_{h_1, \bar{h}_1} \mc{O}_{h_2, \bar{h}_2}: (z , \bar{z} ).
\end{aligned}
\end{equation}
Making a similar assumption about the structure of anti-holomorphic poles in the $ \mc{O}_{h_1, \bar{h}_1} \mc{O}_{h_2, \bar{h}_2}$ OPE, namely that there are no interactions satisfying $s_I =s_1+s_2+1$, we similarly find that 
\begin{equation} \label{Lnbar}
\begin{aligned}
\left[\bar{L}_n, :\mc{O}_{h_1, \bar{h}_1} \mc{O}_{h_2, \bar{h}_2}:(z , \bar{z} )\right] & = \bar{z}^n \left((n+1) (\bar{h}_1+\bar{h}_2) + \bar{z} \bar{\partial} \right): \mc{O}_{h_1, \bar{h}_1} \mc{O}_{h_2, \bar{h}_2}: (z , \bar{z} ).
\end{aligned}
\end{equation}

Our subsequent analysis uses the above form of the transformation of multi-particle operators to determine their coefficients in the multi-particle OPE. Specifically, we constrain terms with \emph{holomorphic} singularities using the \emph{anti-chiral} transformations \eqref{Lnbar}.  Note, our assumption about the absence of interactions with $s_I =s_1+s_2+1$ is an assumption about the structure of the \emph{anti-holomorphic} singularities in the single-particle OPEs.  Hence, a complete analysis that treats these terms must simultaneously deal with both the holomorphic and anti-holomorphic singularity structure of celestial OPEs and is thus beyond the scope of this paper. Nevertheless, it would be interesting and informative to work out the details, especially in light of the known difficulties in working with both holomorphic and anti-holomorphic celestial OPE singularities and given that the correct answer can be found from the Wick method in section \ref{sec:boundary-derivation}. Moreover, as we will see shortly, in our symmetry analysis, the OPE coefficients of single-particle operators and multi-particle operators in the multi-particle OPE are determined independently.  Note however that the corrections arising from certain single-particle OPE channels described above imply that single and multi-particle operators mix under the action of ${\rm SL}(2, \mathbb{C})$ when these channels are present. Thus, it would be interesting to understand in detail how this tightly constrained structure is resolved by the multi-particle celestial OPE. 

In the next subsection we show that translations act non-diagonally on the space of Laurent coefficients in the multi-particle OPE,  similar to the way that translations mix the primaries and descendants of single-particle operators.  Hence, to isolate the constraints from translations on only the primary coefficients in the multi-particle OPE, we need additional transformations that act non-diagonally on multi-particle operators.  The symmetry action generated by $\bar{L}_1$ is such a transformation and in particular we use the following 
\begin{equation}
\begin{aligned}
 \left[\bar{L}_1, : \Big( \bar{\D}^m \mc{O}_{h_1 , \bar{h}_1} \Big)  \mc{O}_{h_2 , \bar{h}_2}:(z , \bar{z} )\right]  
& =  \left(2 (\bar{h}_1 + \bar{h}_2+m) \bar{z} +\bar{z} ^2 \bar{\D}\right) :  \Big(\bar{\D}^m  \mc{O}_{h_1, \bar{h}_1}  \Big)  \mc{O}_{h_2 , \bar{h}_2}: (z , \bar{z} )     \\
\ & \ \ \ + m \left(2 \bar{h}_1 +m-1 \right) : \Big(\bar{\D}^{m - 1}  \mc{O}_{h_1, \bar{h}_1}\Big)   \mc{O}_{h_2 , \bar{h}_2}: (z , \bar{z} )  ,
\end{aligned}
\end{equation}
and 
\begin{equation}
\begin{aligned}
  \left[\bar{L}_1, :\mc{O}_{h_1 , \bar{h}_1}  \bar{\D}^m \mc{O}_{h_2 , \bar{h}_2}:(z , \bar{z} )\right]  
& =   \left(2 (\bar{h}_1 + \bar{h}_2+m)\bar{z} +\bar{z}^2 \bar{\D}\right)  :\mc{O}_{h_1 , \bar{h}_1} \bar{\D}^m \mc{O}_{h_2 , \bar{h}_2}: (z , \bar{z} )   \\ 
& \ \ \ +  m  \left(2 \bar{h}_2+m-1\right): \mc{O}_{h_1 , \bar{h}_1} \bar{\D}^{m - 1}  \mc{O}_{h_2, \bar{h}_2}:(z , \bar{z} )  .
\end{aligned}
\end{equation}
The derivation of these expressions is conceptually similar to but algebraically more involved than the derivation of \eqref{Lnbar}, so here we simply quote the result.  As in \eqref{Ln} and \eqref{Lnbar}, here we assume the absence of single-particle OPE channels that produce single-particle contributions to the action of $L_n$ and $\bar{L}_n$ on multi-particle operators. 

\subsection{Translations of multi-particle operators}
\label{subsec:translations}

We follow a similar strategy as in the previous section to determine the action of translations on multi-particle celestial operators, where again the symmetry transformations can be deduced from the transformation properties of the single-particle operators and the single-particle OPEs.  In particular, the only necessary modification is to replace the stress-tensor $T$ with the appropriate current that generates the bulk translation symmetry.  This current was identified in \cite{Donnay:2018neh}\footnote{More precisely, the graviton current $P_z$ in (4.29) of \cite{Donnay:2018neh} is the anti-holomorphic descendant of $\mathcal{P}$ defined by \eqref{def-graviton-current}. For our analysis it is more natural to work with the primary $\mathcal{P}$ as opposed to its descendant. }  and its operator product expansion with single-particle celestial operators takes the form \cite{Donnay:2018neh,Himwich:2020rro,Puhm:2019zbl,Pate:2019lpp,Banerjee:2020zlg,Guevara:2021abz,Himwich:2021dau} 
\begin{equation}
    \begin{split}
        \mathcal{P}(w, \bar{w}) \mathcal{O}_{h, \bar{h}}(z, \bar{z}) \sim - \frac{1}{2} \frac{\bar{w}-\bar{z}}{w-z}\mathcal{O}_{h+\frac{1}{2}, \bar{h}+\frac{1}{2}}(z, \bar{z}).
    \end{split}
\end{equation}
In theories with propagating gravitons, $\mathcal{P}$ is constructed from a particular mode of a positive helicity graviton 
\begin{equation} \label{def-graviton-current}
    \begin{split}
        \mathcal{P}(z, \bar{z}) = \frac{1}{\kappa} \lim_{\Delta \to 1}(\Delta-1) G^+_{\Delta}(z, \bar{z}). 
    \end{split}
\end{equation}
Note that $\mathcal{P}$ is a primary field of weight $(h, \bar{h}) = \left(\frac{3}{2}, -\frac{1}{2}\right)$.  Then, following \cite{Banerjee:2020zlg,Guevara:2021abz}, we expand $\mathcal{P}$ in anti-holomorphic modes,
\begin{equation} \label{P-mode-exp}
    \begin{split}
        \mathcal{P}(z, \bar{z}) = \sum_{n} \frac{\mathcal{P}_n(z)}{\bar{z}^{-\frac{1}{2}+n}}
    \end{split}
\end{equation}
and extract the anti-chiral half of the action of translations by taking contour integrals: 
\begin{equation} \label{P-mode-action}
    \begin{split}
        \left[P_{-\frac{1}{2}, n}, \mathcal{O}_{h, \bar{h}}(z, \bar{z})\right] & \equiv  (-1)^{\frac{3}{2}+n}  \oint_{z} \frac{dw}{2\pi i} ~\mathcal{P}_n(w)  \mathcal{O}_{h, \bar{h}}(z, \bar{z}) 
          = \frac{1}{2} \bar{z}^{n+ \frac{1}{2}}\mathcal{O}_{h+\frac{1}{2}, \bar{h}+\frac{1}{2}}(z, \bar{z}),
    \end{split}
\end{equation}
where in the above equation $n = \pm \frac{1}{2}$ and the normalization is chosen to reproduce a standard form of the translation action (see for example equations (5.22) and (B.9) in \cite{Himwich:2021dau}). The generators $P_{m,n}$ carry left and right ${\rm SL}(2, \mathbb{C})$ mode labels.  Here, we only use the anti-chiral half of translations to constrain the coefficients of holomorphic singularities in the multi-particle OPE.\footnote{The action of $P_{\frac{1}{2}, \pm \frac{1}{2}}$ can be deduced by working instead with the light-transform of $\mathcal{P}$ \cite{Himwich:2021dau}:
\[ w^{\frac{3}{2}}(z, \bar{z}) \sim \int \frac{d \bar{w}}{2 \pi i } \frac{1}{(\bar{z}- \bar{w})^{3}} \mathcal{P}(z, \bar{w}). \] }

As before, the OPE between the multi-particle operator and the current determines the symmetry transformation of multi-particle operators  
\begin{equation} \label{P-current-composite-OPE}
    \begin{split}
        \mathcal{P}(w, \bar{w}) :\mathcal{O}_{h_1, \bar{h}_1}\mathcal{O}_{h_2, \bar{h}_2}:(z, \bar{z}) \sim -\frac{1}{2} \frac{\bar{w}-\bar{z}}{w-z}  \left(:\mathcal{O}_{h_1+\frac{1}{2}, \bar{h}_1+\frac{1}{2}}\mathcal{O}_{h_2, \bar{h}_2}:(z, \bar{z})+ :\mathcal{O}_{h_1, \bar{h}_1}\mathcal{O}_{h_2+\frac{1}{2}, \bar{h}_2+\frac{1}{2}}:(z, \bar{z}) \right) .
    \end{split}
\end{equation}
Similar to the previous subsection, the expression above assumes that there is no OPE channel $\mathcal{O}_1\mathcal{O}_2 \to \mathcal{O}_I$ with $s_I = s_1+s_2-1$ (corresponding to $p_{12I}=0$) and no OPE channel $\mathcal{O}_1\mathcal{O}_2 \to \mathcal{O}_I$ with $s_I = s_1+s_2+1$ (corresponding to $\bar{p}_{12I}$).  Each of these arise from bulk three-point interactions of scaling dimension $d_V = 4$. While the former channel with $s_I = s_1+s_2-1$ produces double poles in $w-z$ that are ultimately projected out by the contour integral in \eqref{P-mode-action}, the latter corrects the action of the modes of interest $P_{-\frac{1}{2},n}$. As discussed in the previous subsection, incorporating the effect of these terms requires a simultaneous treatment of both the holomorphic and anti-holomorphic singularity structure and is beyond the scope of the current work. 

Thus, focusing on the symmetry transformations that follow from the OPE \eqref{P-current-composite-OPE}, expanding $\mathcal{P}$ in anti-holomorphic modes as in \eqref{P-mode-exp} and taking appropriate contour integrals as in \eqref{P-mode-action}, we find the action of anti-chiral translations on multi-particle operators is given by:
\begin{equation} \label{Pmn-composite}
    \begin{split}
        \left[P_{-\frac{1}{2}, n}, :\mathcal{O}_{h_1, \bar{h}_1}\mathcal{O}_{h_2, \bar{h}_2}:(z, \bar{z}) \right] &\equiv (-1)^{\frac{3}{2}+n}\oint_{z} \frac{dw}{2\pi i} ~\mathcal{P}_n(w)  :\mathcal{O}_{h_1, \bar{h}_1}\mathcal{O}_{h_2, \bar{h}_2}:(z, \bar{z})\\
        & = \frac{ \bar{z}^{n+ \frac{1}{2}}}{2} \left(:\mathcal{O}_{h_1+\frac{1}{2}, \bar{h}_1+\frac{1}{2}}\mathcal{O}_{h_2, \bar{h}_2}:(z, \bar{z})
        +:\mathcal{O}_{h_1, \bar{h}_1}\mathcal{O}_{h_2+\frac{1}{2}, \bar{h}_2+\frac{1}{2}}:(z, \bar{z})\right), 
    \end{split}
\end{equation}
where again $n = \pm \frac{1}{2}$. 

The factor of $\bar{z}$ in the transformation \eqref{Pmn-composite} with $n = +\frac{1}{2}$ indicates that the action of $P_{-\frac{1}{2}, +\frac{1}{2}}$ mixes different orders in a Laurent expansion.  Hence, to study the corresponding constraint, our ansatz must contain an infinite sum over powers of $\bar{z}$, where an appropriate choice of coefficients is derivatives of composites of the form $:(\bar{\partial}^k \mathcal{O}_{1}) \mathcal{O}_{2}:$ and $: \mathcal{O}_{1} (\bar{\partial}^k\mathcal{O}_{2}):$. The action of translations on these composites is
\begin{equation} \label{translation-deriv-1}
    \begin{split}
       & \left[P_{- \frac{1}{2}, n}, : \left(\bar{\D}^k \mc{O}_{h_1, \bar{h}_1} \right)  \mc{O}_{h_2, \bar{h}_2}:(z , \bar{z} ) \right]\\
        &\quad \quad \quad  =\frac{1}{2} \bar{z}^{n+\frac{1}{2}} \left(  :  \left( \bar{\D}^k \mc{O}_{h_1 + \frac{1}{2}, \bar{h}_1 + \frac{1}{2}}\right)   \mc{O}_{h_2, \bar{h}_2} :(z , \bar{z} ) +   :   \left(\bar{\D}^k \mc{O}_{h_1, \bar{h}_1}\right)   \mc{O}_{h_2 + \frac{1}{2}, \bar{h}_2 + \frac{1}{2}}: (z , \bar{z} ) \right) \\
\ &  \quad \quad \quad \quad 
+ \frac{1}{2} k\left(n+ \frac{1}{2}\right)  :  \left(\bar{\D}^{k - 1} \mc{O}_{h_1 + \frac{1}{2}, \bar{h}_1 + \frac{1}{2}} \right)  \mc{O}_{h_2, \bar{h}_2} :(z , \bar{z} ), 
    \end{split}
\end{equation}
and 
\begin{equation} \label{translation-deriv-2}
    \begin{split}
        & \left[P_{- \frac{1}{2}, n}, :  \mc{O}_{h_1, \bar{h}_1}   \bar{\D}^k\mc{O}_{h_2, \bar{h}_2}:(z , \bar{z} ) \right]\\
        &\quad \quad \quad  =\frac{1}{2} \bar{z}^{n+\frac{1}{2}} \left(  :   \mc{O}_{h_1 + \frac{1}{2}, \bar{h}_1 + \frac{1}{2}}   \bar{\D}^k \mc{O}_{h_2, \bar{h}_2} :(z , \bar{z} ) +  :   \mc{O}_{h_1, \bar{h}_1}    \bar{\D}^k\mc{O}_{h_2 + \frac{1}{2}, \bar{h}_2 + \frac{1}{2}}: (z , \bar{z} ) \right) \\
\ &  \quad \quad \quad \quad 
+ \frac{1}{2} k\left(n+ \frac{1}{2}\right)  : \mc{O}_{h_1, \bar{h}_1 }   \bar{\D}^{k - 1}  \mc{O}_{h_2+ \frac{1}{2}, \bar{h}_2 + \frac{1}{2}} :(z , \bar{z} ), 
    \end{split}
\end{equation}
where $n = \pm \frac{1}{2}$. Again, we assume the absence of OPE channels $\mathcal{O}_1\mathcal{O}_2 \to \mathcal{O}_I$ with $s_I = s_1+s_2+1$ discussed above. 

\subsection{Single-particle contributions to the multi-particle OPE from Poincar\'e}
\label{subsec:symmetry-single-particle}  

We now use the Poincar\'{e} transformations found in the previous subsections to constrain the single-particle part of a general composite OPE between celestial primaries.  We begin with the following ansatz for the single-particle contributions to the multi-particle OPE
\begin{equation} \label{ansatz-single-particle-multi-OPE}
    \begin{split}
         \mathcal{O}_{h_1, \bar{h}_1}(z , \bar{z} ) : \mathcal{O}_{h_2, \bar{h}_2} \mathcal{O}_{h_3,\bar{h}_3}:(0,0) 
  \supset \sum_{p \geq 0} \sum_{m=0}^\infty C_p^{(m)}(\bar{h}_1, \bar{h}_2, \bar{h}_3) \frac{\bar{z}^{p+m}}{z^2} \bar{\partial}^m \mathcal{O}_{h_1+h_2+h_3-2, \bar{h}_1+\bar{h}_2+\bar{h}_3+p} (0,0),
    \end{split}
\end{equation}
and determine the constraints from the anti-chiral half of bulk Poincar\'e, \textit{i.e.}~the generators $\bar{L}_m$ and $P_{-\frac{1}{2}, n}$. As in the single-particle OPE analysis, the sum over $p$ is a proxy for the sum over the spin of the produced operator, which here is equal to $s_J = s_1+s_2+s_3-2-p$.  Also, like in the single-particle OPE analysis, we include the infinite sum over anti-holomorphic descendants because we will use the anti-chiral translation generator $P_{-\frac{1}{2},\frac{1}{2}}$, which acts non-diagonally on the space of single-particle primaries and anti-holomorphic descendants.  However, unlike the single-particle OPE analysis, here we specify a double-pole in the holomorphic coordinate.  Although we already know this to be the correct form from previous sections, this ansatz could nevertheless be motivated by the observation that single-particle contributions to the multi-particle OPE arise when a \emph{pair} of propagators go on-shell and thus produce a \emph{double} pole.\footnote{It would be interesting if the form of the multi-particle OPE ansatz in this and the following subsections could be derived from symmetry, for example, by generalizing the analyses in \cite{Banerjee:2020zlg,Banerjee:2020vnt,Banerjee:2025oyu}. }  We do not consider the simple-pole single-particle contributions to the multi-particle OPE that were found in subsection \ref{subsec:general-ope} and arise from overlapping holomorphic and anti-holomorphic singularities. It is possible that their derivation from symmetry is related to the modified symmetry transformations in the presence of $\mathcal{O}_1\mathcal{O}_2 \to \mathcal{O}_I$ interactions with $s_I = s_1+s_2+1$ that were discussed in the previous subsections. 

We now use the symmetry transformation properties from previous subsections to determine the coefficients $C^{(m)}_p$. Applying $\left[ \bar{L}_1, \cdot \right]$ to the left-hand side of \eqref{ansatz-single-particle-multi-OPE} and isolating the resulting single-particle contribution, we find 
\begin{equation}
\begin{aligned}
  &[ \bar{L}_1, \mc{O}_{h_1, \bar{h}_1}(z, \bar{z}) :\mc{O}_{h_2, \bar{h}_2} \mc{O}_{h_3, \bar{h}_3}:(0, 0)] \\
&  \quad \supset   \sum_{p \geq 0}\sum_{m=0}^\infty (2 \bar{h}_1 +  p + m) \frac{\bar{z}^{ p + m + 1}}{z^2} C_p^{(m)}(\bar{h}_1, \bar{h}_2, \bar{h}_3)  \bar{\D}^m \mc{O}_{h_1 + h_2 + h_3 - 2, \bar{h}_1 + \bar{h}_2 + \bar{h}_3 +  p}(0,0),
\end{aligned}
\end{equation} 
while on the right-hand side, we have 
\begin{equation}
\begin{aligned}
 & \Big[ \bar{L}_1,  \sum_{p \geq 0}\sum_{m=0}^\infty \frac{\bar{z}^{p + m}}{z^2} C_p^{(m)}(\bar{h}_1, \bar{h}_2, \bar{h}_3)  \bar{\D}^m \mc{O}_{h_1 + h_2 + h_3 - 2, \bar{h}_1 + \bar{h}_2 + \bar{h}_3 + p}(0,0) \Big] \\
&  \quad =  \sum_{p \geq 0}\sum_{m=0}^\infty \frac{\bar{z}^{p + m + 1}}{z^2} C_p^{(m+1)}(\bar{h}_1, \bar{h}_2, \bar{h}_3)  (m{+}1) (2 \bar{h}_1 {+} 2 \bar{h}_2 {+} 2 \bar{h}_3 {+} 2p {+} m) \bar{\D}^m   \mc{O}_{h_1 + h_2 + h_3 - 2, \bar{h}_1 + \bar{h}_2 + \bar{h}_3 + p} (0,0).
\end{aligned}
\end{equation} 
Equating the two gives the recursion relation 
\begin{equation}\label{singleparticlelorentz}
(2 \bar{h}_1 +  p + m)  C_p^{(m)}(\bar{h}_1, \bar{h}_2, \bar{h}_3)  =  (m+1) (2 \bar{h}_1 + 2 \bar{h}_2 + 2 \bar{h}_3 + 2p + m) C_p^{(m+1)}(\bar{h}_1, \bar{h}_2, \bar{h}_3).   
\end{equation} 
Instead applying $[P_{- \frac{1}{2}, \pm \frac{1}{2}}, \cdot ]$, we find the respective new constraints \begin{equation}\label{singleparticletranslation}
\begin{aligned}
C^{(m)}_p \LP \bar{h}_1 {+} \frac{1}{2}, \bar{h}_2, \bar{h}_3 \RP & = (m+1) C_p^{(m+1)}(\bar{h}_1, \bar{h}_2, \bar{h}_3), \\
C^{(m)}_p(\bar{h}_1, \bar{h}_2, \bar{h}_3) & = C_p^{(m)} \LP \bar{h}_1 {+} \frac{1}{2}, \bar{h}_2, \bar{h}_3 \RP + C_p^{(m)} \LP \bar{h}_1, \bar{h}_2 {+} \frac{1}{2}, \bar{h}_3 \RP +  C_p^{(m)} \LP \bar{h}_1, \bar{h}_2, \bar{h}_3 {+} \frac{1}{2} \RP.
\end{aligned}
\end{equation}
Taking appropriate linear combinations then gives the fixed-$m$ initial condition 
\begin{equation}  \label{single-particle-constraint-sys}
\begin{aligned}
(2 \bar{h}_1 {+}  p)  C_p^{(0)}(\bar{h}_1, \bar{h}_2, \bar{h}_3)  & = 2( \bar{h}_1 {+}  \bar{h}_2 {+}  \bar{h}_3 {+} p) C_p^{(0)}\LP \bar{h}_1 {+} \frac{1}{2}, \bar{h}_2, \bar{h}_3 \RP, \\
( 2 \bar{h}_2 {+} 2 \bar{h}_3 {+}  p) C_p^{(0)}\LP \bar{h}_1, \bar{h}_2, \bar{h}_3 \RP & = 2( \bar{h}_1 {+}  \bar{h}_2 {+}  \bar{h}_3 {+} p) \left[ C_p^{(0)}\LP \bar{h}_1, \bar{h}_2 {+} \frac{1}{2}, \bar{h}_3 \RP {+} C_p^{(0)}\LP \bar{h}_1, \bar{h}_2, \bar{h}_3 {+} \frac{1}{2} \RP\right].
\end{aligned}
\end{equation} 
This is analogous to the system of constraints  
\begin{equation}
\begin{aligned}
(2 \bar{h}_1 + p)  C_p^{(0)}(\bar{h}_1, \bar{h}_2) & = 2 (\bar{h}_1 + \bar{h}_2 + p) C_p^{(0)} \LP \bar{h}_1 + \frac{1}{2}, \bar{h}_2 \RP, \\
(2 \bar{h}_2 + p) C_p^{(0)}(\bar{h}_1, \bar{h}_2) & = 2( \bar{h}_1 + \bar{h}_2 + p) C_p^{(0)} \LP \bar{h}_1, \bar{h}_2 + \frac{1}{2} \RP,
\end{aligned}
\end{equation}
found in \cite{Pate:2019lpp} for the corresponding single-particle OPE, but since we have an additional variable $\bar{h}_3$, the system is under-constrained.  In appendix \ref{app:singleParticleRecursion} we solve the system of constraints in \eqref{single-particle-constraint-sys} and find the general solution  
\begin{equation} \label{int-sol-p'}
    \begin{split}
         B(2\bar{h}_1 +p, 2 \bar{h}_2+p', 2 \bar{h}_3+p-p'),
    \end{split}
\end{equation}
where $p'$ is a free parameter. Thus, the general $m=0$ coefficient is given by an arbitrary linear combination of the solution \eqref{int-sol-p'}
\begin{equation} \label{final-sol-4.3}
    \begin{split}
        C^{(0)}_p (\bar{h}_1, \bar{h}_2, \bar{h}_3) = \sum_{p'} \alpha_{pp'} B(2\bar{h}_1 +p, 2 \bar{h}_2+p', 2 \bar{h}_3+p-p').
    \end{split}
\end{equation}
Then, identifying
\begin{equation}
    p = p_{12I}+p_{I3J}= s_1+s_2+s_3-s_J-2  , \quad \quad p' = p_{12I} = s_1+s_2-s_I-1,
\end{equation}
we find that \eqref{final-sol-4.3} precisely matches the single-particle contribution to the result \eqref{generalOPE-primary-holomorphic}, where   the sum over $p'$ becomes a sum over $I$,  the sum over $p$ becomes the sum over $J$ and $\alpha_{pp'} = \gamma^{s_1, s_2}_{s_I} \gamma^{s_I, s_3}_{s_J}$.  Finally, the OPE coefficients for the single-particle descendants in \eqref{generalOPE-holomorphic} satisfy the recursion in $m$ in \eqref{singleparticlelorentz}.  

\subsection{Multi-particle contributions to the multi-particle OPE from Poincar\'e}
\label{subsec:symmetry-multi-particle}

The same logic extends to the multi-particle contributions to the multi-particle OPE,  where again the symmetry-based derivation begins with an ansatz.  Here we consider an ansatz of the form  
\begin{equation} \label{multi-multi-ansatz}
    \begin{split}
        \mathcal{O}_{h_1, \bar{h}_1}(z , \bar{z} ) &: \mathcal{O}_{h_2, \bar{h}_2} \mathcal{O}_{h_3,\bar{h}_3}:(0,0) \\
 & \supset \sum_{p \geq 0} \sum_{m=0}^\infty D_p^{(m)}(\bar{h}_1, \bar{h}_2, \bar{h}_3) \frac{\bar{z}^{p+m}}{z} : \left(\bar{\partial}^m \mathcal{O}_{h_1+h_2-1, \bar{h}_1+\bar{h}_2+p} \right) \mathcal{O}_{h_3, \bar{h}_3}: (0,0)\\
  &  \quad \quad  +\sum_{p \geq 0} \sum_{m=0}^\infty E_p^{(m)}(\bar{h}_1, \bar{h}_2, \bar{h}_3) \frac{\bar{z}^{p+m}}{z} :\mathcal{O}_{h_2, \bar{h}_2}\bar{\partial}^m \mathcal{O}_{h_1+h_3-1, \bar{h}_1+\bar{h}_3+p}: (0,0),
    \end{split}
\end{equation}
and we constrain the coefficients with the anti-chiral bulk Poincar\'e generators $\bar{L}_m$ and $P_{-\frac{1}{2}, n}$. In this ansatz, we assume that the singularity structure takes the form of simple poles in $z$, but otherwise allow for general (non-singular) anti-holomorphic structure. We also assume that the composites appearing on the right-hand side are formed from a pair of single-particle operators where one of the pair has the same conformal dimensions as one of the single-particle operators in the composite on the left-hand side. This assumption is motivated by the results of the previous sections, for example \eqref{generalOPE-holomorphic}, but it would be interesting to find an independent intrinsic justification, perhaps from associativity or crossing symmetry. Like in the other ans\"{a}tze, the sum over $p$ is equivalent to a sum over the conformal spin of the operator whose conformal dimensions depend on $p$. In the terms involving $:(\bar{\partial}^m \mathcal{O}_I )\mathcal{O}_3:$, $p$ is related to the conformal spin by $s_I = s_1+s_2-1-p$ and in the terms involving $: \mathcal{O}_2\bar{\partial}^m \mathcal{O}_I:$  by $s_I = s_1+s_3-1-p$. Finally, similar to the previous ans\"{a}tze \eqref{single-single-ansatz} and \eqref{ansatz-single-particle-multi-OPE}, we include an infinite sum over powers of $\bar{z}$ since bulk translations mix different terms in a Laurent expansion. Curiously, we find that the requisite terms are not exactly global conformal descendants of the composites $\bar{\partial}^m:\mathcal{O}_i \mathcal{O}_j:$, but rather composites with a specified derivative structure,  $:(\bar{\partial}^m \mathcal{O}_{i} )\mathcal{O}_{j}:$ or $: \mathcal{O}_{i} \bar{\partial}^m\mathcal{O}_{j}:$,  which transform in a relatively simple way under translations as in \eqref{translation-deriv-1} and \eqref{translation-deriv-2}.

Applying $[ \bar{L}_1, \cdot]$ to both sides of the ansatz \eqref{multi-multi-ansatz}, we find on the left-hand side 
\begin{equation}
\begin{aligned}
 &[\bar{L}_1, \mc{O}_{h_1, \bar{h}_1} (z, \bar{z}): \mc{O}_{h_2, \bar{h}_2} \mc{O}_{h_3, \bar{h}_3} : (0,0)] \\
\ &  \quad \supset \sum_{p\geq 0} \sum_{m=0}^\infty \LP 2 \bar{h}_1 + m + p \RP \frac{\bar{z}^{m+p+1}}{z}  D_p^{(m)}(\bar{h}_1, \bar{h}_2, \bar{h}_3) : \left(  \bar{\D}^m \mc{O}_{h_1 + h_2 - 1, \bar{h}_1 + \bar{h}_2 + p}  \right) \mc{O}_{h_3, \bar{h}_3}: (0,0) \\
\ & \ \ \ \quad + \sum_{p\geq 0} \sum_{m=0}^\infty \LP 2 \bar{h}_1 + m + p \RP \frac{\bar{z}^{m+p+1}}{z} E_p^{(m)}(\bar{h}_1, \bar{h}_2, \bar{h}_3) :\mc{O}_{h_2, \bar{h}_2} \bar{\D}^m \mc{O}_{h_1 + h_3 - 1, \bar{h}_1 + \bar{h}_3 + p} : (0,0),
\end{aligned}
\end{equation} 
and on the right-hand side
\begin{equation}
\begin{aligned}
& \Bigg[ \bar{L}_1, \sum_{p\geq 0}\sum_{m=0}^\infty \frac{\bar{z}^{m+p}}{z}  D_p^{(m)}(\bar{h}_1, \bar{h}_2, \bar{h}_3)  :  \left( \bar{\D}^m \mc{O}_{h_1 + h_2 - 1, \bar{h}_1 + \bar{h}_2 + p}\right)  \mc{O}_{h_3, \bar{h}_3}: (0,0) \\
\ & \ \ \ + \sum_{p\geq 0}\sum_{m=0}^\infty \frac{\bar{z}^{m+p}}{z} E_p^{(m)}(\bar{h}_1, \bar{h}_2, \bar{h}_3) :\mc{O}_{h_2, \bar{h}_2} \bar{\D}^m \mc{O}_{h_1 + h_3 - 1, \bar{h}_1 + \bar{h}_3 + p} : (0,0) \Bigg] \\
& = \sum_{p\geq 0}\sum_{m=0}^\infty \frac{\bar{z}^{m+p+1}}{z}  D_p^{(m+1)}(\bar{h}_1, \bar{h}_2, \bar{h}_3) (m+1) (2(\bar{h}_1 + \bar{h}_2 + p) + m) :  \left( \bar{\D}^m  \mc{O}_{h_1 + h_2 - 1, \bar{h}_1 + \bar{h}_2 + p} \right)  \mc{O}_{h_3 , \bar{h}_3}: (0, 0) \\
\ & \ \ \ +\sum_{p\geq 0} \sum_{m=0}^\infty \frac{\bar{z}^{m+p+1}}{z} E_p^{(m+1)}(\bar{h}_1, \bar{h}_2, \bar{h}_3) (m+1)(2 (\bar{h}_1 + \bar{h}_3 + p) + m) : \mc{O}_{h_2 , \bar{h}_2} \bar{\D}^m  \mc{O}_{h_1 + h_3 - 1, \bar{h}_1 + \bar{h}_3 + p}:(0, 0).
\end{aligned}
\end{equation} 
Here we have kept only the multi-particle terms that appear in the general transformation of a multi-particle operator under bulk Poincar\'e.  As explained in subsection \ref{subsec:sl2c} and \ref{subsec:translations}, this is equivalent to assuming the absence of single-particle OPE channels $IJ \to K$ with $s_K = s_I+s_J+1$, where here $(s_I, s_J) = (s_1+s_2-1-p, s_3)$ and  $(s_I, s_J) = ( s_2,s_1+s_3-1-p)$.  The presence of such channels implies relations between the OPE coefficients of single-particle and multi-particle contributions and may, for example, play a critical role in reproducing the subleading single-particle contributions derived in subsection \ref{subsec:general-ope} from overlapping holomorphic and anti-holomorphic singularities. Such an analysis is left to future work. 

${\rm SL}(2, \mathbb{C})$ invariance is enforced by equating the two results, which gives the following two constraints: 
\begin{equation}
\begin{aligned}
\left(2 \bar{h}_1 +p+ m \right) D_p^{(m)}(\bar{h}_1, \bar{h}_2, \bar{h}_3) & = (m+1) \left(2 (\bar{h}_1 + \bar{h}_2 + p)  +  m \right) D_p^{(m+1)}(\bar{h}_1, \bar{h}_2, \bar{h}_3) , \\
\left(2 \bar{h}_1+p + m  \right)  E_p^{(m)}(\bar{h}_1, \bar{h}_2, \bar{h}_3) & = (m+1) \left(2 (\bar{h}_1 + \bar{h}_3 + p) + m\right) E_p^{(m+1)}(\bar{h}_1, \bar{h}_2, \bar{h}_3).
\end{aligned}
\end{equation} 
Covariance with respect to the translation generators $[P_{- \frac{1}{2}, \pm \frac{1}{2}}, \cdot ]$ similarly gives 
\begin{equation}
\begin{aligned}
D_p^{(m)} \LP \bar{h}_1 + \frac{1}{2}, \bar{h}_2, \bar{h}_3 \RP & = (m+1) D_p^{(m+1)} \LP \bar{h}_1, \bar{h}_2, \bar{h}_3 \RP, \\
D_p^{(m)} \LP \bar{h}_1 + \frac{1}{2}, \bar{h}_2, \bar{h}_3 \RP  + D_p^{(m)} \LP \bar{h}_1, \bar{h}_2 + \frac{1}{2}, \bar{h}_3 \RP & =  D_p^{(m)} \LP \bar{h}_1, \bar{h}_2, \bar{h}_3 \RP ,\\
D_p^{(m)} \LP \bar{h}_1, \bar{h}_2, \bar{h}_3 + \frac{1}{2} \RP & = D_p^{(m)} \LP \bar{h}_1, \bar{h}_2, \bar{h}_3 \RP ,\\
E_p^{(m)} \LP \bar{h}_1 + \frac{1}{2}, \bar{h}_2, \bar{h}_3 \RP & = (m+1) E_p^{(m+1)} \LP \bar{h}_1, \bar{h}_2, \bar{h}_3 \RP, \\
E_p^{(m)} \LP \bar{h}_1 + \frac{1}{2}, \bar{h}_2, \bar{h}_3 \RP + E_p^{(m)} \LP \bar{h}_1, \bar{h}_2, \bar{h}_3 + \frac{1}{2} \RP & = E_p^{(m)} \LP \bar{h}_1, \bar{h}_2, \bar{h}_3 \RP, \\
E_p^{(m)} \LP \bar{h}_1, \bar{h}_2 + \frac{1}{2}, \bar{h}_3 \RP & = E_p^{(m)} \LP \bar{h}_1, \bar{h}_2, \bar{h}_3 \RP,
\end{aligned}
\end{equation} 
where again we have assumed the absence of interactions that modify the symmetry transformations \eqref{Pmn-composite}, \eqref{translation-deriv-1} and \eqref{translation-deriv-2}. Note that the constraints on $D_p^{(m)}$ mirror those on $E_p^{(m)}$, with only the roles of $\bar{h}_2, \bar{h}_3$ swapped.  

Taking appropriate linear combinations at $m = 0$ gives the simplified set of constraints for $D_p^{(0)}$  
 \begin{equation}
\begin{aligned}
(2 \bar{h}_1 + p  ) D_p^{(0)}(\bar{h}_1, \bar{h}_2, \bar{h}_3) & = 2 (\bar{h}_1 + \bar{h}_2 + p) D_p^{(0)} \LP \bar{h}_1 + \frac{1}{2}, \bar{h}_2, \bar{h}_3 \RP, \\
(2 \bar{h}_2 + p) D_p^{(0)}(\bar{h}_1, \bar{h}_2, \bar{h}_3) & = 2 (\bar{h}_1 + \bar{h}_2 + p) D_p^{(0)} \LP \bar{h}_1 , \bar{h}_2 + \frac{1}{2}, \bar{h}_3 \RP, \\
D_p^{(0)} \LP \bar{h}_1, \bar{h}_2, \bar{h}_3 + \frac{1}{2} \RP & = D_p^{(0)} \LP \bar{h}_1, \bar{h}_2, \bar{h}_3 \RP.
\end{aligned}
\end{equation}
Under mild analytic assumptions (convexity on the real line suffices, for example), the third constraint implies $D^{(0)}_p$ is independent of $\bar{h}_3$. Then the first two constraints reduce to a simpler system, equivalent to the non-composite case studied by \cite{Pate:2019lpp}. From that previous analysis, the unique solution at $m = 0$ is  
\begin{equation}
D^{(0)}_p(\bar{h}_1, \bar{h}_2, \bar{h}_3) = \gamma^{s_1, s_2}_{s_I} B(2 \bar{h}_1 + p, 2 \bar{h}_2 + p), \quad \quad s_I = s_1+s_2-p-1,
\end{equation}
or after applying the recursion relations in $m$ 
\begin{equation}
D_{p}^{(m)}(\bar{h}_1, \bar{h}_2, \bar{h}_3) = \frac{\gamma^{s_1, s_2}_{s_I} }{m!} B( 2 \bar{h}_1+p + m, 2 \bar{h}_2+p ).
\end{equation}
An identical analysis for $E_{p}^{(m)}$ gives 
\begin{equation}
E_{p}^{(m)}(\bar{h}_1, \bar{h}_2, \bar{h}_3) = \frac{\gamma^{s_1, s_3}_{s_I}  }{m!} B(2 \bar{h}_1+p+ m,2 \bar{h}_3+p), \quad \quad \quad s_I = s_1+s_3-p-1.
\end{equation}
Again we find that the multi-particle OPE coefficients implied by the anti-chiral half of bulk Poincar\'{e} symmetry precisely match the result for the holomorphic singular terms found in \eqref{generalOPE-holomorphic}.  

\section{Discussion} \label{sec:discussion}

\subsection*{Summary}

In this work, we present three complementary methods for determining the multi-particle celestial operator product expansions and demonstrate precise agreement between the results in a variety of different contexts.  The first method involving the generalized Wick formula is generally superior, since it can be justified on pure mathematical grounds, without any assumption about the nature or existence of a celestial conformal field theory.  Nevertheless, to further understand the structure of bulk collinear limits, it is instructive to see in detail the precise way to recover the same result directly from scattering amplitudes.  From the symmetry perspective, it is also informative to establish that the Wick method is bulk-translation covariant. In particular, since bulk translational symmetry is obscured in the conformal primary basis, it is not a priori obvious that 2D conformally covariant techniques will necessarily produce bulk-translationally covariant results. Our analysis in section \ref{sec:symmetry} establishes that the multi-particle OPE indeed represents a bulk-translation covariant decomposition of a celestial amplitude into lower-point amplitudes. Thus, all together, the three different methods provide multiple overlapping perspectives on the tightly-constrained structure of the underlying multi-particle OPE and in particular, each can be viewed as a different entry point for a bottom-up/bootstrap approach to celestial holography. 
 
\subsection*{Restricted set of elementary celestial primary fields}
 
Our results provide compelling evidence that multi-particle celestial operators are not \emph{new} primary fields in celestial conformal field theory. Rather, multi-particle celestial operators can be treated as familiar composite operators in conformal field theory, whose OPE coefficients are determined by the OPE coefficients of the constituents.  It is still an open question whether the single-particle celestial operators constitute the complete set of primaries in celestial conformal field theory.  Nevertheless, evidence against the proliferation of additional independent states (specifically specified by independent data) representing arbitrarily many particles is an important and encouraging sign for the overall tractability of the bottom-up approach to celestial holography. 

\subsection*{Implications for the consistent implementation of universal enveloping algebras}

This work also has essential consequences for the consistent implementation of the universal enveloping algebra of the known holographic symmetry algebras in gauge theory and gravity.  In a standard two-dimensional conformal field theory, symmetry transformations generated by composites of the stress tensor $:TT \cdots T:$ can be re-expressed by contour integral deformation in terms of elements of the universal enveloping algebra of the Virasoro algebra, namely linear combinations of successive application of symmetry transformations generated by a single stress tensor, $L_{n_1} \cdots L_{n_m}$ where $L_n = \oint \frac{dz}{2 \pi i } z^{n+1} T(z)$. The coefficients of OPEs involving $:TT:$ are thus tightly constrained to reproduce the correct action of the corresponding generators of the universal enveloping algebra. Hence, from this perspective, it is crucial that the OPEs involving $:TT:$ are determined by OPEs involving $T$, since the former generates elements of the universal enveloping algebra of the symmetry generated by the latter.  

Celestial conformal field theories are known to admit the much larger symmetry algebras of ${\rm w}_{1+\infty}$ and the `S'-algebra  \cite{Guevara:2021abz,Strominger:2021lvk,Himwich:2021dau}, generated by (contour integrals of) conformally soft gluons and gravitons, respectively.  These symmetry algebras are encoded in the delicate boost-weight $\Delta_i$ pole structure of generic single-particle celestial OPE coefficients involving gluons and gravitons.  It is further expected that states in celestial conformal field theory transform in representations of the universal enveloping algebra of ${\rm w}_{1+\infty}$ and the `S'-algebra, for which operator product expansions involving composites of the associated currents become relevant.  Similar to the stress tensor and Virasoro symmetry in standard 2D CFT, we expect contour integrals of composites of the ${\rm w}_{1+\infty}$ and the `S'-algebra currents to generate elements of the corresponding universal enveloping algebras.  This in turn places stringent constraints on coefficients of OPEs involving the composite currents or equivalently, on the boost-weight pole structure of the multi-gluon and graviton celestial OPE coefficients. The analysis herein establishes that the multi-particle OPEs inherit their pole structure from that of the single-particle OPEs. It is beyond the scope of this work to confirm whether the Wick formula derivation automatically enforces all of the necessary constraints on the multi-particle OPE, but we expect this to be the case. 

Exciting prospects for celestial holography arise from novel constraints on celestial amplitudes or the corresponding momentum-space scattering amplitudes.  Constraints in the form of differential equations are particularly powerful and a number of these have already been identified \cite{Banerjee:2020zlg,Banerjee:2020vnt,Banerjee:2021cly,Banerjee:2021dlm,Banerjee:2023zip,Ruzziconi:2024zkr,Hu:2021lrx,Banerjee:2023bni,Banerjee:2025grp}.  In many of these investigations, soft current algebra descendants play a key role in formulating null states and deriving the associated differential constraints \cite{ Banerjee:2020zlg,Banerjee:2020vnt,Banerjee:2021cly,Banerjee:2021dlm,Banerjee:2023zip,Hu:2021lrx,Banerjee:2023bni,Banerjee:2025grp,Ruzziconi:2024zkr}. More generally, current algebra descendants were found to play a key role in the formulation and organization of an all-orders celestial operator product expansion in the MHV sector of gauge theory and gravity \cite{Ebert:2020nqf, Adamo:2022wjo, Ren:2023trv}.  The techniques developed herein, specifically those pertaining to the implementation of the universal enveloping algebra, may open new avenues for the discovery of additional constraints implied by the soft symmetry algebras.
 
\subsection*{Loop corrections and associativity}

While all explicit expressions for celestial amplitudes and OPEs in this paper strictly  pertain to tree-level scattering in asymptotically-flat spacetimes, we expect the generalized Wick method of section \ref{sec:boundary-derivation} and symmetry-based method of section \ref{sec:symmetry} to extend to loop-level. As emphasized above, the Wick method is guaranteed by the mathematics of complex analysis, and  the treatment of branch cuts seems to be the only potential subtlety that might arise in a loop-level analysis.  As for the symmetry method, loop-level amplitudes ultimately respect bulk Poincar\'e symmetry and so this symmetry should ultimately manifest in the appropriately corrected OPEs for these amplitudes. 

The associativity of the multi-particle OPEs is also an interesting question, which we leave to a future investigation. The associativity of the single-gluon contribution to the all-negative-helicity multi-gluon OPEs was previously investigated in \cite{Guevara:2024ixn}. It would be interesting both to check the associativity of the multi-particle celestial OPE derived herein, as well as to extend the analysis to loop-level and study the effect of the one-loop corrections associated to anomaly-cancellation in twistor space \cite{Costello:2022wso,Costello:2022upu,Bittleston:2022jeq,Fernandez:2023abp,Fernandez:2024qnu,Serrani:2025oaw}, as well as more general higher-loop multi-particle corrections \cite{Zeng:2023qqp}.  In particular, a central result of this paper is the OPE coefficients of multi-particle contributions to the multi-particle celestial OPE, which have not yet been fully determined by previous investigations. While conditions like the double-residue condition \cite{Ren:2022sws} and the corresponding three-particle factorization channel statement \cite{Ball:2022bgg} give insight into the associativity properties of the single-particle contributions to the multi-particle OPE \cite{Guevara:2024ixn}, it is not clear that these are the relevant conditions for the multi-particle contributions. Likely, the most simple approach would be to check associativity (or the Jacobi identity) directly, but after, it would be interesting to determine the analogous requisite conditions on scattering amplitudes.  At least in the conformally soft limit, associativity of the multi-particle OPE is likely related to the consistent implementation of the universal enveloping algebra discussed above.  Finally, it would be interesting and instructive to investigate associativity in the presence of the additional simple pole single-particle contributions to the multi-particle OPE derived in subsection \ref{subsec:general-ope}.  In particular, these terms contain information about both holomorphic and anti-holomorphic singularities, while associativity analyses to date have primarily focused on only one or the other. 

\subsection*{Subleading collinear limits of scattering amplitudes}

Finally, our results and the boundary method presented in section \ref{sec:boundary-derivation} provide new insight into the structure of scattering  amplitudes at subleading orders in the collinear limit.  Such limits may be of general interest to the amplitudes community.  For example, they play an integral role in the striking relation between scattering of gluons and gravitons discovered in \cite{Stieberger:2015kia}.\footnote{We thank Tomasz Taylor for bringing this connection to our attention.}

\section*{Acknowledgments} 

We are grateful to Mina Himwich, Lukas Lindwasser, Panos Oikonomou, Sruthi Narayanan, Atul Sharma, David Skinner, Andy Strominger and Tomasz Taylor for insightful conversations and comments during the course of this project. We thank Alfredo Guevara and Mina Himwich for comments on a draft.  Finally, we are deeply grateful to Marcus Hoskins for his general contribution to the final stages of this project, and especially for thoughtful and detailed feedback on multiple drafts. This work was completed with the support of NSF grant 2310633.  

\begin{appendix}

\section{Mathematics underlying the Wick formula} \label{app:explain-Wick}

Some of the formal structure of 2D CFTs reduces to that of complex functions of several variables. In this appendix we demonstrate that the generalized Wick theorem \ref{generalizedwick-1} is an example of this paradigm. 

Consider a function of the form
\begin{equation}
f(z_1, z_2, z_3) = \frac{1}{z_{12}^A z_{13}^B z_{23}^C} \mbox{ for } A,B,C \in \mathbb{Z}_{>0}.
\end{equation}
Here we show that the result of ``normal-ordering in $23$'', $i.e.$ extracting the $\Theta(z_{23}^0)$ term in the limit $z_2 \to z_3$, and then extracting the singular terms in the limit $z_1 \to z_3$ is exactly reproduced by the procedure specified by the Wick formula \eqref{generalizedwick-1}.

First, we use 
\begin{equation}
    f(z_1, z_2, z_3)
        = \frac{1}{z_{13}^{A+B}z_{23}^C} \sum_{k=0}^\infty  {A+k-1 \choose k} \left(\frac{z_{23}}{z_{13}}\right)^k
\end{equation}
to normal-order $f$ in $23$ as described above:
\begin{equation} \label{f-normal-23}
    \begin{split}
        f(z_1, z_2, z_3) \Big|_{\Theta(z_{23}^0)}
            ={A+C-1 \choose C}\frac{1}{z_{13}^{A+B+C}}.
    \end{split}
\end{equation}
Then, this is exactly the singular term in the limit $z_1 \to z_3$. 

On the other hand, to apply the generalized Wick method, we first need to extract the singular terms in the limit $z_1 \to z_2$ and $z_1\to z_3$.  We find these are given respectively by
\begin{equation}
    \begin{split}
        f(z_1, z_2, z_3)\Big|_{\text{singular in }z_{12}} &= \frac{1}{z_{12}^A z_{23}^{B+C}} \sum_{k=0}^{A} (-1)^k {B+k-1 \choose k} \left(\frac{z_{12}}{z_{23}}\right)^k  , \\
        f(z_1, z_2, z_3)\Big|_{\text{singular in }z_{13}} &= \frac{(-1)^A}{z_{13}^B z_{23}^{A+C}} \sum_{k=0}^{B} {A+k-1 \choose k} \left(\frac{z_{13}}{z_{23}}\right)^k .
    \end{split}
\end{equation}
Finally, the content of the generalized Wick theorem is the statement that the following contour integral of the above limits reproduces the singular terms  in $f(z_1, z_2, z_3) \big|_{\Theta(z_{23}^0)}$:
\begin{equation}
    \begin{split}
       f(z_1, z_2, z_3) \Big|_{\Theta(z_{23}^0)} \Big|_{\text{singular in }z_{13}} =\oint_{z_3} \frac{dz_2}{2 \pi i} \frac{1}{z_{23}} \left[f(z_1, z_2, z_3)\Big|_{\text{singular in }z_{12}} + f(z_1, z_2, z_3)\Big|_{\text{singular in }z_{13}}\right].
    \end{split}
\end{equation} 
Evaluating this explicitly, we find 
\begin{equation}
    \begin{split}
        \oint_{z_3}& \frac{dz_2}{2 \pi i} \frac{1}{z_{23}} \left[f(z_1, z_2, z_3)\Big|_{\text{singular in }z_{12}} + f(z_1, z_2, z_3)\Big|_{\text{singular in }z_{13}}\right]
        \\
        &=\oint_{z_3} \frac{dz_2}{2 \pi i} \frac{1}{z_{23}} \left[\frac{1}{z_{12}^A z_{23}^{B+C}} \sum_{k=0}^{A} (-1)^k {B+k-1 \choose k} \left(\frac{z_{12}}{z_{23}}\right)^k  + \frac{(-1)^A}{z_{13}^B z_{23}^{A+C}} \sum_{k=0}^{B} {A+k-1 \choose k} \left(\frac{z_{13}}{z_{23}}\right)^k \right]\\
        &=\oint_{z_3} \frac{dz_2}{2 \pi i} \frac{1}{z_{23}}  \sum_{k=0}^{A} (-1)^k {B+k-1 \choose k} \frac{1}{z_{23}^{B+C+k}} \sum_{j=0}^\infty  {A-k+j-1 \choose j} 
        \frac{z_{23}^j}{z_{13}^{A-k+j}} \\
        &= \frac{1}{z_{13}^{A+B+C}} \sum_{k=0}^{A} (-1)^{k} {B+k-1 \choose k} {A+B+C-1 \choose B+C+k}  \\
        &= {A+C-1 \choose C}\frac{1}{z_{13}^{A+B+C}},
    \end{split}
\end{equation}
which precisely matches the right-hand side of \eqref{f-normal-23}.

\section{Spinor helicity conventions} \label{sec:conventions}

In the bulk we use similar conventions as in \cite{Pasterski:2017ylz}. We work in the mostly-plus metric \begin{equation}
\eta_{\mu \nu} = \begin{bmatrix}
-1 & 0 & 0 & 0 \\
0 & 1 & 0 & 0 \\
0 & 0 & 1 & 0 \\
0 & 0 & 0 & 1
\end{bmatrix}.
\end{equation}
We parametrize null momenta in (3+1)-dimensional Minkowski space as
\begin{equation}
p^\mu(\omega, z, \bar{z}) = \epsilon \omega (1 + z \bar{z}, z + \bar{z}, -i(z - \bar{z}), 1 - z \bar{z}) \equiv \epsilon \omega \hat{p}^\mu(z, \bar{z}),
\end{equation}
where $\omega > 0$ and $\epsilon = \pm 1$ for outgoing and incoming states, respectively. Their inner product is given by
\begin{equation}
p_1 \cdot p_2 \equiv \eta_{\mu \nu} p^\mu (\omega_1, z_1, \bar{z}_1) p^\nu(\omega_2, z_2, \bar{z}_2) = - 2 \epsilon_1 \epsilon_2 \omega_1 \omega_2 z_{12} \bar{z}_{12}.
\end{equation} 
A null momentum can be written equivalently in spinor-helicity notation as a $2 \times 2$ non-invertible matrix $p_{a \dot{a}} = \tilde{\lambda}_a \lambda_{\dot{a}}$, whose constituent Weyl spinors are parametrized as follows \begin{equation}\label{spinorparametrization}
\begin{bmatrix}
\lambda_{\dot{1}} \\ \lambda_{\dot{2}} 
\end{bmatrix} = \epsilon \sqrt{2 \omega} \begin{bmatrix}
z \\ -1 
\end{bmatrix} \implies \begin{bmatrix}
\lambda^{\dot{1}} \\ \lambda^{\dot{2}}
\end{bmatrix} = \epsilon \sqrt{2 \omega} \begin{bmatrix} -1 \\ -z \end{bmatrix} \equiv \lambda(\omega, z)
\end{equation}
and \begin{equation}
\begin{bmatrix}
\tilde{\lambda}_1 \\ \tilde{\lambda}_2 \end{bmatrix} = \sqrt{2 \omega} \begin{bmatrix}- \bar{z} \\ 1 \end{bmatrix} \implies \begin{bmatrix}
\tilde{\lambda}^1 \\ \tilde{\lambda}^2
\end{bmatrix}  = \sqrt{2 \omega} \begin{bmatrix} 1 \\ \bar{z} \end{bmatrix} \equiv \tilde{\lambda}(\omega, \bar{z}).
\end{equation}
Here we raise and lower indices with the Levi-Civita symbol
\begin{equation}
\chi^a = \varepsilon^{ab} \chi_b \mbox{ and } \chi_a = \varepsilon_{ab} \chi^b,
\end{equation}
where\begin{equation}
\varepsilon^{12} = - \varepsilon^{21} = 1 \mbox{ and } \varepsilon_{12} = - \varepsilon_{21} = -1.
\end{equation}
This gives \begin{equation}
p_{a \dot{a}} = p_\mu \sigma^\mu_{a \dot{a}} = 2 \epsilon \omega \begin{bmatrix}
- z \bar{z} & \bar{z} \\
z & - 1 
\end{bmatrix}
\end{equation}
where \begin{equation}
(\sigma^\mu) = (\mathbf{1}, \vec{\sigma}).
\end{equation}
These spinors have their own inner products \begin{equation}\label{bracketvalues}
\begin{aligned}
\langle 12 \rangle & \equiv \varepsilon^{\dot{a}\dot{b}} (\lambda_1)_{\dot{a}} (\lambda_2)_{\dot{b}} = - 2 \epsilon_1 \epsilon_2 \sqrt{\omega_1 \omega_2} z_{12} = - \langle 21 \rangle, \\
[ 12 ] & \equiv - \varepsilon^{ab} (\tilde{\lambda}_1)_a (\tilde{\lambda}_2)_b = 2 \sqrt{\omega_1 \omega_2} \bar{z}_{12} = - [21],
\end{aligned}
\end{equation}
related to the momentum vector inner product by \begin{equation}
\langle 12 \rangle [ 12 ] = 2 p_1 \cdot p_2.
\end{equation}

\section{Solution for single-particle constraint relations}\label{app:singleParticleRecursion}

In this appendix we solve the system of equations in \eqref{single-particle-constraint-sys}, generalizing the method outlined in appendix E of \cite{Pate:2019lpp}. For ease of notation, we first denote \begin{equation}
\begin{aligned}
(x,y,z) \equiv (2 \bar{h}_1 + p, 2 \bar{h}_2, 2 \bar{h}_3 + p)
\end{aligned}
\end{equation}
and rewrite $C_p^{(0)}$ as \begin{equation}
C_p^{(0)} \LP \bar{h}_1, \bar{h}_2, \bar{h}_3 \RP = C_p^{(0)} \LP \frac{x - p}{2} , \frac{y}{2}, \frac{z - p}{2} \RP \equiv F(x,y,z),
\end{equation}
where $p$ is treated as a background constant rather than an argument. Then \eqref{single-particle-constraint-sys} becomes \begin{equation}\label{Fconstraints}
\begin{aligned}
xF(x,y,z) & = (x+y+z)F(x+1,y,z), \\
(y+z)F(x,y,z) & = (x+y+z)[F(x,y+1,z) + F(x,y,z+1)].
\end{aligned}
\end{equation}
Summing both sides and removing common terms gives the constraint \begin{equation}
F(x,y,z) = F(x+1,y,z) + F(x,y+1,z) + F(x,y,z+1),
\end{equation}
which is solved in particular by the trivariate Euler beta function. Inspired by, this we rewrite  \begin{equation}
F(x,y,z) = B(x,y,z) \Phi(x,y,z)
\end{equation}
and solve equivalently for $\Phi$ satisfying \begin{equation}
\begin{aligned}
\Phi(x+1,y,z) & = \Phi(x,y,z), \\
(y+z)\Phi(x,y,z) & = y\Phi(x,y+1,z) + z \Phi(x,y,z+1).
\end{aligned}
\end{equation}
This system of difference equations is linear and homogeneous with a vector space of solutions.  We can therefore expand in some convenient basis of ans\"{a}tze and constrain those. We use the familiar ansatz of factorizable functions \begin{equation}
\Phi(x,y,z) = \Phi_1(x) \Phi_2(y) \Phi_3(z).
\end{equation}
The first constraint reduces to the statement that $\Phi_1$ is period-one \begin{equation}
\Phi_1(x+1) = \Phi_1(x),
\end{equation}
which has basis solutions \begin{equation}
\Phi_1(x) = e^{i 2 \pi n_x x}
\end{equation}
for $n_x \in \mathbb{Z}$. The second constraint becomes the functional equation \begin{equation}
(y+z)\Phi_2(y) \Phi_3(z) = y \Phi_2 (y+1) \Phi_3 (z) + z\Phi_2(y) \Phi_3(z+1)
\end{equation}
or equivalently \begin{equation}
z \LP 1 -   \frac{\Phi_3(z+1)}{\Phi_3(z)} \RP = - y \LP 1 - \frac{\Phi_2(y+1)}{\Phi_2(y)} \RP = p'
\end{equation}
for $p'$ an undetermined constant. These single-variable difference equations have basis solutions \begin{equation}
\Phi_2(y) = \frac{\Gamma(y + p')}{\Gamma(y)}e^{i 2 \pi n_y y}, \ \ \ \ \ \ \Phi_3(z)  = \frac{\Gamma(z - p')}{\Gamma(z)} e^{i 2 \pi n_z z},
\end{equation}
for $n_y, n_z \in \mathbb{Z}$. Thus, our original system  has a solution space spanned by the functions
\begin{equation}
\begin{aligned}
F_{\vec{n}} (x,y,z) & = B(x,y,z) \frac{\Gamma(y + p')}{\Gamma(y)} \frac{\Gamma(z - p')}{\Gamma(z)} e^{i 2 \pi  \vec{n} \cdot \vec{x}} \\
& = B(x,y + p', z - p') e^{i 2 \pi \vec{n} \cdot \vec{x}}
\end{aligned}
\end{equation}
for $\vec{n} \in \mathbb{Z}^3$. We can naturally restrict to the subspace $\vec{n} = 0$ by demanding, \textit{e.g.}, convexity in each argument. Returning to our original notation and expanding in this basis, we have equivalently \begin{equation} 
C^{(0)}_p \LP \bar{h}_1, \bar{h}_2, \bar{h}_3 \RP   = \sum_{p'} \alpha_{pp'}  B(2 \bar{h}_1 + p, 2 \bar{h}_2 + p', 2 \bar{h}_3 + p - p'). 
\end{equation}
As noted in subsection \ref{subsec:symmetry-single-particle}, the constants $p,p'$ admit a direct physical interpretation in terms of the dimensions $p_{12I}, p_{I3J}$ of the bulk interaction vertices mediating the OPE channels that compose to give single-particle contributions to composite OPEs.

\end{appendix}

\bibliography{multiparticle_celestial_ope}
\bibliographystyle{utphys}

\end{document}